\documentclass[bibyear]{aa} 

\usepackage{graphicx}
\usepackage{booktabs}
\usepackage{txfonts}
\usepackage{amsmath}
\usepackage{wasysym}
\usepackage{siunitx}
\usepackage{float}
\usepackage{natbib}

\usepackage[colorlinks=true,citecolor=blue,linkcolor=blue]{hyperref}
\AtBeginDocument{}

\usepackage{xcolor}
\usepackage{ulem}
	
\usepackage{lineno}

\begin{document}

   \title{Deep XMM-Newton observation reveals extraplanar emission in NGC 5746}

\titlerunning{Extraplanar emission in NGC 5746}

   \author{Roman Laktionov
          \and 
          Manami Sasaki
          \and 
          J\"orn Wilms
          }

\authorrunning{Laktionov et al.}

   \institute{Dr.\ Karl Remeis-Observatory, Erlangen Centre for Astroparticle Physics, Friedrich-Alexander-Universität Erlangen-Nürnberg, Sternwartstraße 7, 96049 Bamberg, Germany
              \email{rom.laktionov@fau.de}
             } 
   \date{Received 1 October 2025; accepted 16 March 2026}

  \abstract
   {We present a deep \textit{XMM-Newton} observation of the massive edge-on galaxy NGC~5746. The total exposure time of ${\sim}250\,\mathrm{ks}$ provides unprecedented sensitivity to study the diffuse hot gas in the halo, significantly surpassing the depth of previous observations.}
   {While the presence of hot, circumgalactic gas has been thoroughly tested in starburst galaxies, detections in normal galaxies remain scarce. By studying the diffuse X-ray emission in NGC~5746, we aim to constrain the extent and physical properties of its halo.}
   {We created X-ray images and surface-brightness profiles to quantify the distribution of extraplanar gas in NGC~5746. In addition, we isolated the diffuse emission from point-source and background contamination and studied its spectral characteristics.}
   {We detected diffuse extraplanar emission in the azimuthally averaged surface-brightness profile out to a projected radius of ${\sim}30\,\mathrm{kpc}$ with an integrated significance of $3\sigma$. The emission is anisotropic: The western quadrant shows a significant excess at ${\sim}1.5\mbox{--}3.5\arcmin$ (integrated $3.7\sigma$), while the eastern quadrant exhibits a marginal enhancement in a single radial bin. The northern and southern quadrants are consistent with the background within uncertainties.
   The X-ray luminosity of the hot halo is ${\sim}8.8\cdot10^{39}\,\mathrm{erg\,s^{-1}}$ in 0.3--2.0\,keV. Spectral modeling yielded a plasma temperature of ${\sim}0.6\,\mathrm{keV}$, while the disk emission is dominated by nonthermal emission from unresolved X-ray sources.}
   {Our results provide evidence of the presence of diffuse hot gas in the halo of NGC~5746 and suggest that earlier nondetections may be attributed to sensitivity limits. The asymmetric morphology highlights the complexity of the halo emission. Distinguishing between the possible origin scenarios for the extraplanar gas requires future observations with improved sensitivity and spectral resolution.}

   \keywords{X-rays: galaxies, ISM
               }

   \maketitle

\section{Introduction}

The interaction of galaxies with the circumgalactic medium plays a key role in the study of galactic evolution. Feedback from massive stars can produce a blowout of hot gas in the form of chimneys \citep{norman1989}, galactic fountains \citep{shapiro1976, bregman1980, avillez2000}, or winds \citep{breitschwerdt1991, breitschwerdt1999}. Young stars create cavities in the interstellar medium (ISM) through radiation pressure and stellar winds. Shock waves from supernova (SN) explosions insert energy into these cavities, causing them to expand and create bubbles of hot, low-density gas. If enough energy is injected, and the bubbles become sufficiently large, they can break out of the galactic disk and form an outflow.

In starburst galaxies, the combined energy input from massive stars can drive superwinds with mass outflow rates up to 10--20 times the star formation rate (SFR) of the host galaxy \citep{hopkins2012}. The expelled gas reaches temperatures of $T > 10^6\,\mathrm{K}$ \citep{mcquinn2018}, creating X-ray plumes \citep{schulz1992, pietsch2000, strickland2000, pietsch2001} and supershells \citep{heiles1984} that can be traced at soft X-ray energies, mainly below ${\sim}1\,\mathrm{keV}$. This hot ionized medium likely originates from swept-up and shock-heated halo gas or from compressed SN ejecta at the edge of the outflow \citep{strickland2004b}.

Empirically, the distribution of the hot ionized medium correlates well with diffuse ionized gas that has been photoionized by OB-stars \citep{reynolds1984, tüllmann2000, wood2004, strickland2004b, tüllmann2006b} and blown out of the disk by SN explosions. This diffuse gas can be traced by H$\alpha$ emission that extends 1.5--10\,kpc into the circumgalactic medium and provides evidence of the disk-halo interaction.

In addition to X-rays and H$\alpha$ emission, the halos of starburst galaxies can also be traced by radio continuum emission. It is produced via synchrotron radiation of high-energy cosmic rays (CRs) interacting with the magnetic field of the halo. Since the CRs require evolved SN remnants (SNRs) to be accelerated in significant numbers, the CR-halo is formed after the X-ray and diffuse ionized gas halo \citep{tüllmann2006a}.

While the multifrequency correlation of the X-ray and H$\alpha$ components is well established for starburst galaxies \citep{condon1992, wang2001, read2001, ranalli2003, strickland2004b, tüllmann2006b}, for normal star-forming galaxies, the existence of such a correlation remains a matter of debate. On the one hand, a blowout of hot, ionized gas is a rare occurrence in normal galaxies, as it requires a very massive star cluster that is powerful enough to penetrate the disk \citep{ferrara2000a, ferrara2000b}. \citet{tüllmann2006b} found that the X-ray luminosity of the halo, $L_\mathrm{X,halo}$, is correlated with the SFR in the disk. On the other hand, X-ray studies of normal edge-on galaxies with clearly separable disk and halo components are still scarce. 
In their \textit{Chandra} and \textit{XMM-Newton} studies of extraplanar X-ray emission of star-forming galaxies, \citet{strickland2004b} and \citet{tüllmann2006a} found that most galaxies with detected X-ray halos are starbursts, but they also found that the X-ray emission of the normal galaxy NGC~891 is consistent with the SFR-$L_\mathrm{X,halo}$ correlation. Similarly, the \textit{Chandra} data of NGC~5746 also indicate the presence of an X-ray halo in this normal star-forming galaxy \citep{pedersen2006}.

NGC~5746 is a barred spiral galaxy, classified as SAB(rs)b in the Third Reference Catalog of Bright Galaxies \citep{devaucoleurs1991}. It is viewed nearly edge-on \citep[$i = 86\fdg{}8$][]{bianchi2007}) and is relatively massive, with a stellar mass of $M_\ast = 1.0\mbox{--} 1.3\cdot10^{11}\,\mathrm{M_{\odot}}$ \citep{jiang2019}. Its circular velocity is also rather high, with $v_\mathrm{circ} = (311 \pm 8)\,\mathrm{km\,s^{-1}}$ in the outer disk \citep[HyperLEDA][]{hyperleda}. We adopted a distance of $(29 \pm 5)\,\mathrm{Mpc}$ (HECATE; \citealt{hecate}), where $1'$ on the sky corresponds to ${\sim}8.4\,\mathrm{kpc}$. The most important parameters of NGC~5746 are listed in Table~\ref{main_par}.

The galaxy was extensively studied in the near-infrared and optical bands \citep{barentine2012, kormendy2019, martig2021}, in X-rays \citep{pedersen2006, rasmussen2009}, and in radio \citep{rand2008}. It is the largest member of the NGC~5746 group of galaxies \citep{garcia1993} and forms a very wide galaxy pair with NGC~5740 \citep{martig2021}. Despite the presence of a large extension of gas to the northeast of NGC 5740, possibly caused by ram pressure stripping \citep{rand2008}, there is consensus in the literature that NGC~5746 does not show signs of recent interaction events \citep{pedersen2006, barentine2012, martig2021}. \citet{rand2008} deem a tidal stripping scenario unlikely due to the companion's morphology and large distance to NGC~5746 of ${\sim}155\,\mathrm{kpc}$.

\citet{martig2021} studied the formation history of NGC~5746 and found that the galaxy's bar and extended disk, containing ${\sim}80\%$ of the stellar mass, formed more than $10\,\mathrm{Gyrs}$ ago. An inflow of gas along the bar toward the center of the galaxy led to the formation of a nuclear disk and a boxy/peanut-shaped bulge via vertical thickening of the bar, which extends ${\sim}35''$ along the major axis. \citeauthor{martig2021} detected a metal-poor stellar population in the thick disk, whose age distribution indicates that the only significant merger happened ${\sim}8\,\mathrm{Gyrs}$ ago. This merger contributed roughly 30\% of the accreted stars in the disk while leaving the galaxy's morphology unchanged. Other than that, NGC~5746 has had a quiescent history dominated by minor mergers and a bar-driven secular evolution \citep{kormendy2019}. Around the bar, a bright ring of gas ($r {\sim}1'$) with elevated levels of star formation formed, while the activity inside the ring remained low. Although the disk shows signs of star formation in the last 4--5\,Gyr \citep{martig2021}, 
at a level of $0.8\mbox{--}1.0\,\mathrm{M_{\odot}\,yr^{-1}}$ \citep{jiang2019}, the current rate of star formation is far below the levels expected for a starburst galaxy.

The low level of star formation in NGC~5746 raises interesting questions regarding the galaxy's evolution, as \citet{pedersen2006} reported the discovery of a hot X-ray halo extending out to at least $20\,\mathrm{kpc}$ from the galactic center when using a $36.8\,\mathrm{ks}$ observation with the \textit{Chandra} Advanced CCD Imaging Spectrometer (ACIS) from April 2003. The discrepancy lies in the fact that SFR levels below  ${\sim}1\,\mathrm{M_{\odot}\,yr^{-1}}$ are generally considered insufficient to generate the SN rates necessary to produce a blow-out of hot X-ray emitting gas into the environment \citep{tüllmann2006b}. \citet{pedersen2006} estimated the necessary energy input to produce the observed emission to be on the order of $E_\mathrm{min} {\sim} 10^{57}\mbox{--}10^{58}\,\mathrm{erg}$. Such a release of energy must have involved a strong burst of star formation in the past, but it is unclear where such a starburst could have come from, given the quiescent merger history of NGC~5746. In addition, \citet{rasmussen2009} failed to detect any significant, extraplanar H$\alpha$ or radio continuum emission expected for a stellar outflow. \citet{pedersen2006} concluded that the most viable explanation for NGC~5746's hot halo is ongoing accretion of primordial gas from the intergalactic medium (IGM), as predicted by cosmological models \citep{benson2000, toft2002}. They found that the X-ray properties of low-metallicity gas falling onto the galaxy at $10-20\,\mathrm{km\,s^{-1}}$ are consistent with their observation. Despite their claim, the authors later retracted the detection of a hot X-ray halo around NGC~5746. In a more recent study, they applied newer calibration data in their analysis and found that the halo properties are consistent with spatial variations in the quantum efficiencies of the ACIS charge-coupled devices (CCDs) \citep{rasmussen2009}.

In this paper, based on four deep \textit{XMM-Newton} observations, we show that an X-ray halo is indeed present around NGC~5746. It is visible in both X-ray imaging (Fig.~\ref{xmm_img}) and in surface brightness profiles (Fig.~\ref{brightness_profiles}). Based on spectroscopic evidence from these observations, we show that the X-ray halo may have been produced by stellar feedback from present or past star formation activity, rather than accretion of primordial gas from the IGM. 
In Sect.~\ref{chap2} we outline the peculiarities of the camera on board \textit{XMM-Newton} and the reduction of our X-ray data. In Sect.~\ref{chap3} we describe the details of the image analysis and the spectral modeling. In Sects.~\ref{chap4} and~\ref{chap5} we present our results and compare them to previous studies. Our conclusions regarding the nature of the outflow are presented in Sect.~\ref{chap6}.

\begin{table}
    \centering
    \caption{Properties of NGC~5746.}\label{main_par}
    \begin{tabular}{p{5cm} l}
\hline       $\alpha_\mathrm{J2000}$ & 14h 44m 55.89s   \\
             $\delta_\mathrm{J2000}$ & $1^{\circ}\,57\arcmin\,18\farcs0$ \\
             $D_\mathrm{25}\tablefootmark{a}$ & $(7.24 \pm 0.11)\,\mathrm{arcmin}$ \\
             $v_\mathrm{circ}\tablefootmark{a}$ & $(311 \pm 8)\,\mathrm{km/s}$\\
             Distance\tablefootmark{b} & $(29 \pm 5)\,\mathrm{Mpc}$ \\
             Inclination\tablefootmark{c} & $86\fdg8$\\
             Stellar mass\tablefootmark{d} & $(1.1-1.3)\cdot10^{11}\,\mathrm{M_{\odot}}$\\
             SFR\tablefootmark{d} & $(0.8-1.0)\,\mathrm{M_{\odot}\,yr^{-1}}$\\\hline
    \end{tabular}
\tablefoot{
\tablefoottext{a}{\citet{hyperleda}}
\tablefoottext{b}{\citet{hecate}}
\tablefoottext{c}{\citet{bianchi2007}}
\tablefoottext{d}{\citet{jiang2019}}}
\end{table}

\section{Data}\label{chap2}

Due to the larger effective area and better low-energy response of the European Photon Imaging Camera (EPIC) compared to the \textit{Chandra} ACIS, \textit{XMM-Newton} is more sensitive at energies below ${\sim} 1\,\mathrm{keV}$ compared to \textit{Chandra} and thus better suited for the search of faint, diffuse emission in the halo of galaxies. In addition, the larger field of view (FOV) of \textit{XMM-Newton} compared to \textit{Chandra} allows us to extend the search to greater distances from the center of the galaxy and to extract a background spectrum that is less contaminated by halo emission from the source. We selected our target galaxy based on a search in the XMM-Newton science archive\footnote{\url{https://www.cosmos.esa.int/web/xmm-newton/xsa}} for unpublished X-ray data of star-forming galaxies. We found a set of four deep observations of the spiral galaxy NGC~5746 (PI: Jesper Rasmussen). The observations were conducted within a time span of one month, between 2011 January 10 and 2011 February 10. The observations had a duration of 80--90\,ks each, with a total exposure time of 341\,ks (see Table~\ref{tab_exptimes} for the filtered exposure times).

    \begin{table}
        \centering
        \caption{Observation properties.}\label{tab_exptimes}
        \begin{tabular}{lcccc}
          \hline\hline                              
           ObsID & MOS1 [ks] & MOS2 [ks] & PN [ks]  \\ 
           \hline
           0651890101 & 55.1 & 58.3 & 42.7 \\
           0651890201 & 37.2 & 40.5 & 29.8 \\
           0651890301 & 85.6 & 86.7 & 76.3 \\
           0651890401 & 70.0 & 74.7 & 56.6 \\
           \hline
        \end{tabular}
\tablefoot{Observation IDs (ObsIDs) and exposure times of the EPIC-MOS1, -MOS2, and -PN detectors after filtering the time intervals contaminated by soft proton flares.}
    \end{table}

\subsection{XMM-Newton}

The EPIC camera on board the \textit{XMM-Newton} spacecraft is composed of three detectors: two metal-oxide-semiconductor \citep[MOS][]{mos} CCD arrays and one pn \citep{pn} CCD camera. Each camera is positioned in the focal plane of a $7.5\,\mathrm{m}$ Wolter I grazing-incidence X-ray telescope, of which two carry a reflection grating array (RGA). The RGAs divert about half of the incoming flux toward reflection grating spectrometers (RGS). The EPIC-MOS cameras are positioned behind the telescopes that contain the RGAs and receive ${\sim}44\%$ of the original flux, whereas the EPIC-pn camera is located behind the third telescope and is unaffected by reflection gratings. The RGS are designed for high-resolution spectral studies of point sources, whereas the EPIC cameras have a larger effective area and are better suited for studies of faint, extended sources. The EPIC cameras have a spectral resolution of $E/\Delta E {\sim}20$--50 in the 0.15--15\,keV energy range, as well as very sensitive imaging capabilities with an angular resolution of $6\,\text{arcsec}$ at the full width at half maximum of a point source. The large FOV of $28\farcm4$ in diameter permits a detailed image analysis of large, diffuse extragalactic sources.

\subsection{Data processing}\label{sect_processing}

The data were processed using the XMM-Newton Extended Source Analysis Software\footnote{\url{https://www.cosmos.esa.int/web/xmm-newton/sas}} (ESAS), version~1.3. We first used the \texttt{emanom} task to examine the MOS CCDs for anomalous states based on hardness ratio calculations for each chip. MOS1-4 was in an anomalous state during all observations, while MOS1-6 was inactive due to micrometeorite strikes. MOS2-2 was in an anomalous state during the first observation and exhibited relatively low hardness ratios in the remaining observations as well. We excluded it from all four observations to ensure consistency between the datasets. The remaining CCDs have been classified as good in all but two (intermediate) cases.

    \begin{table}
        \centering 
        \caption{Number of detected point sources in each observation.}\label{tab_ps}
        \begin{tabular}{lccc}
          \hline\hline                               Obs.ID & MOS1  & MOS2  & PN   \\\hline
           0651890101 & 31 & 48 & 175 \\
           0651890201 & 28 & 36 & 129 \\
           0651890301 & 40 & 64 & 231 \\
           0651890401 & 39 & 56 & 205 \\\hline
        \end{tabular}
    \end{table}

We reduced the data for the MOS and PN observations with the \texttt{emchain} and \texttt{pnchain} tasks. All observations were carried out in full frame mode with the medium optical blocking filter. Time intervals that are significantly contaminated by soft proton (SP) flares were filtered via the \texttt{espfilt} task, which identifies contaminated time intervals by comparing FOV light curves to the corner data. The filtered exposure times are listed in Table~\ref{tab_exptimes}. In total, the effective exposure time is $248\,\mathrm{ks}$ for the MOS1, $260\,\mathrm{ks}$ for the MOS2, and $205\,\mathrm{ks}$ for the PN detector. Although this procedure removes obviously contaminated time intervals, it does not necessarily remove all SP effects. The residual emission is discussed in Sect.~\ref{chap3}.

As we are interested in the diffuse emission produced by hot, ionized gas, we removed the contamination from point sources in the images and spectra. We generated a point source catalog with the \texttt{edetect$\_$chain} algorithm and then used the \texttt{cheese} task to create source masks for all point sources with a minimum detection likelihood $\geq 15$. The number of excluded point sources per observation is listed in Table~\ref{tab_ps}.

The extraction of point source-removed images and spectra was done via the \texttt{mosspectra}/\texttt{pnspectra} tasks. We extracted these products for the soft (0.3--0.7\,keV), medium (0.7--1.2\,keV), and hard (1.2--5\,keV) energy bands, taking care to remove bad and hot pixels. For the PN we extracted single events only, in order to reduce the electronic noise at soft energies. For the MOS we extracted single, double, and quadruple events.

\section{X-ray analysis}\label{chap3}

\subsection{Image}\label{sect_images}

\begin{figure*}
    \centering
    \includegraphics[width=0.49\linewidth]{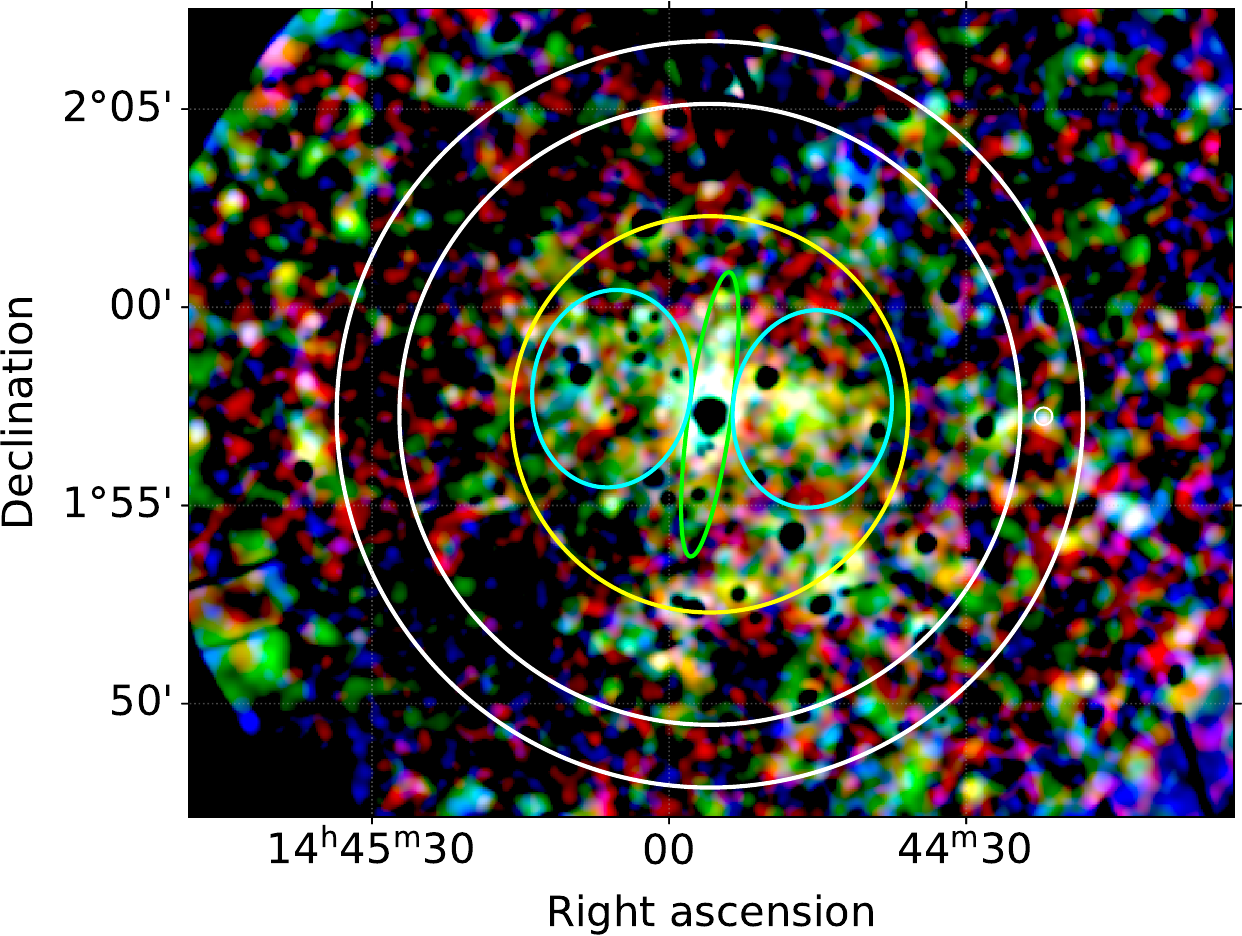}
    \includegraphics[width=0.49\linewidth]{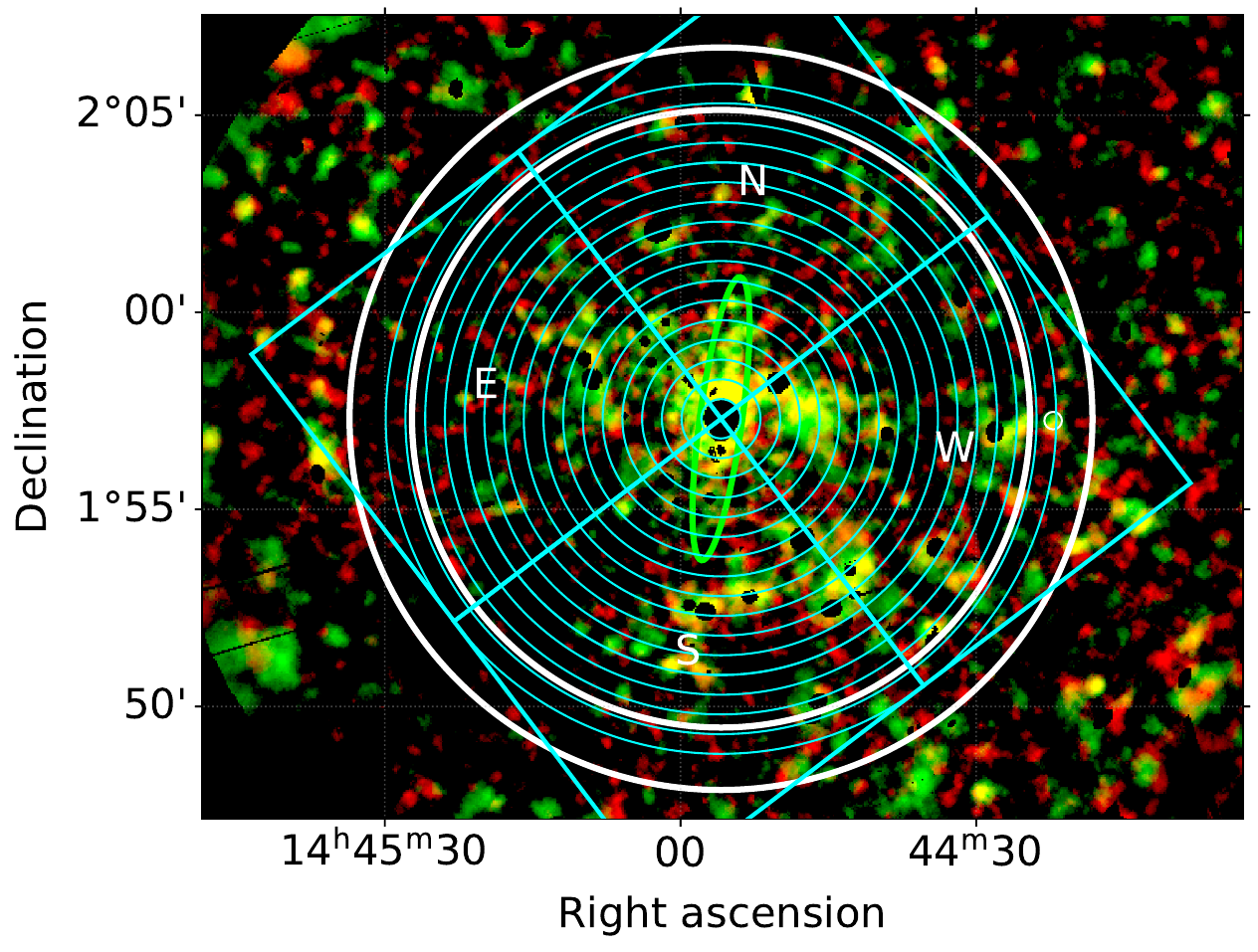}
    \caption{Left: Merged three-color EPIC X-ray image of the four NGC~5746 observations. Red colors represent the soft X-ray band (0.3--0.7\,keV), green colors represent the medium 0.7--1.2\,keV band, and blue colors represent the hard 1.2--5.0\,keV band. The image is displayed using a 60--99 percentile intensity stretch. The green ellipse displays the $D_{25}$ ellipse of NGC~5746. The cyan ellipses display the two regions of enhanced emission, the yellow circle represents the halo ($r = 5\arcmin$), the white annulus displays the background region ($r=7\farcm8\mbox{-}9\farcm4$). Right: Image of the same field in 0.3--1.2\,keV , displayed using a 75-99 percentile stretch. Sixteen circular regions are shown, divided into four quadrants (east, north, west, south) for the construction of the surface brightness profiles.}\label{xmm_img} 
\end{figure*}

The construction of the X-ray image requires further steps involving the ESAS routine and spectral fitting. We aimed to produce a merged three-color image of all four observations to enhance the visibility of faint emission outside the galactic disk. Following the data processing steps outlined in Sect.~\ref{sect_processing}, we combined the filtered event files of each observation using the merge task.

Images and spectra of the quiescent particle background (PB) were generated using the \texttt{mosback}/\texttt{pnback} tasks. In addition, as discussed above, the images are contaminated by residual SP emission, as well as contributions from solar wind charge exchange (SWCX). Determining the level of SP and SWCX contamination requires fitting of the spectral data in the entire FOV. All spectral fitting was done with XSPEC\footnote{\url{https://heasarc.gsfc.nasa.gov/xanadu/XSPEC/}}. We used 0.3--11\,keV data for the MOS detectors, while excluding the 4--6\,keV and 7.2--9.2\,keV data for the PN detector due to the presence of instrumental lines, such as the Ti K$\alpha$ line at 4.51\,keV and the Cr K$\alpha$ line at 5.41\,keV. Furthermore, the observations are contaminated by Al- and Si-K$\alpha$ lines at 1.46 and 1.739\,keV, respectively. The spectral model therefore requires the inclusion of an astrophysical background, an instrumental background, a particle background, and a SWCX background component. A detailed description of the spectral model is provided in Sect.~\ref{sect_bkg}.

The bulk of the PB was easily removed from the images using the PB spectra generated by \texttt{mosback}/\texttt{pnback}. To determine the level of residual SP and SWCX emission, we fitted our model to the PB-subtracted spectra of the merged observations. The \texttt{proton} task calculates the number of residual protons and generates SP images using the power-law index and normalization of the PB spectral model. The SWCX emission was modeled by multiple Gaussian lines. We found a small SWCX contamination from the $0.65\,\mathrm{keV}$ \ion{O}{viii}, $0.81\,\mathrm{keV}$ \ion{O}{viii}, $0.92\,\mathrm{keV}$ \ion{Ne}{ix}, and $1.35\,\mathrm{keV}$ \ion{Mg}{xi} lines. The \texttt{swcx} task determines the number of SWCX photons and generates SWCX images using the normalizations of the Gaussian lines.

The \texttt{combimage} task combines the MOS1, MOS2, and PN images and scales the datasets to the response of the MOS2 medium filter assuming a power-law spectrum with photon index $\alpha = 1.7$. The PB, SP, and SWCX images were also processed during this step. Finally, we used the \texttt{binadapt} task to create exposure-corrected images with an adaptive smoothing kernel of 50. The resulting images of the soft, medium, and hard band were then combined to a smoothed three-color image using the python package astropy\footnote{\url{https://www.astropy.org/}}.

The final product is shown in Fig.~\ref{xmm_img} (left panel). The central ellipse displays the $D_{25}$ angular extent of NGC 5746 (from HECATE). The green ellipses mark two hand-drawn regions with enhanced X-ray emission (East and West). The spectra of the galactic disk and the halo regions are analyzed in Sect.~\ref{sect_spectra}. The yellow circle has a radius of $5\,\text{arcmin}$ (corresponding to ${\sim}42\,\mathrm{kpc}$). We refer to the region inside this circle as the halo. The red annulus marks the region used to extract the background spectrum at a distance of $7\farcm8\mbox{--}9\farcm4$ from the center of the galaxy. One point source has been excluded from the background region by hand, as it is relatively faint and has not been detected by the source detection algorithm. We compared the merged image to the nonmerged images of the four observations and found no discrepancies in the observed trends.

\subsection{Surface brightness profiles}

\begin{figure*}
    \centering
    \includegraphics[width=0.49\linewidth]{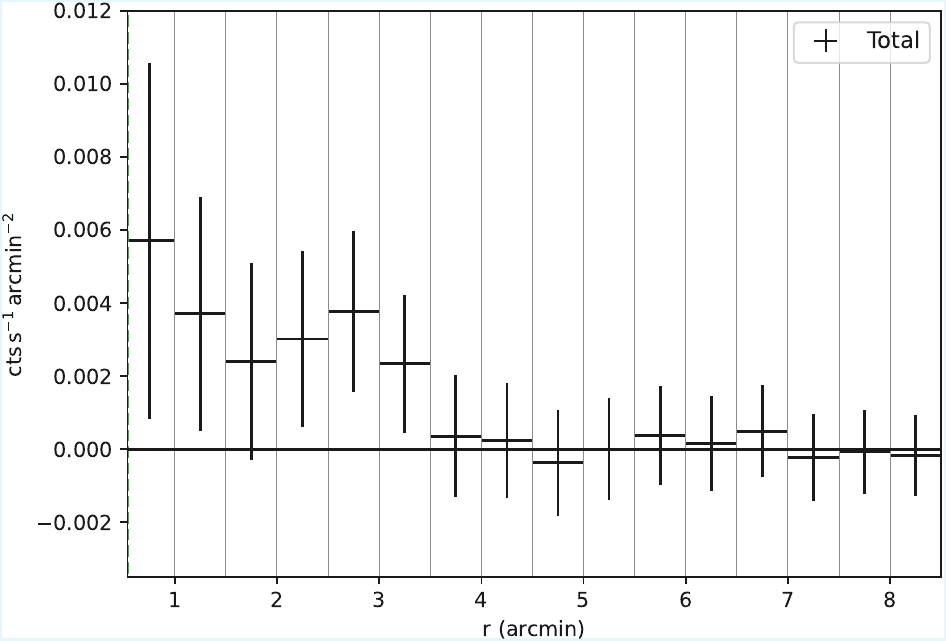}
    \includegraphics[width=0.49\linewidth]{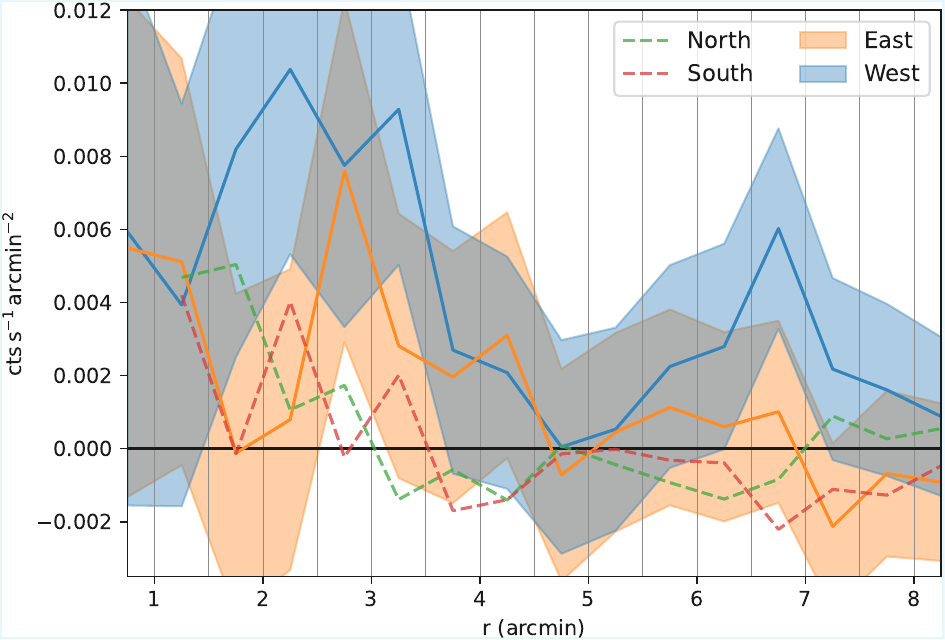}
    \caption{EPIC-pn radial surface brightness profile of the extraplanar emission in NGC~5746. The left panel shows the azimuthally averaged profile, while the right panel presents the profiles in the four quadrants (north, west, south, east). The shaded regions indicate the total $1\sigma$ uncertainties, including both Poisson errors and an additional background variance term derived from ${\sim} 1.8\,\mathrm{arcmin}^2$ subregions and scaled with area. The horizontal line marks zero net surface brightness after background subtraction.}\label{brightness_profiles}
\end{figure*}

In order to quantify the level of diffuse X-ray emission outside of the galactic disk, we created radial surface brightness profiles of the extraplanar emission. This was done for (a) full annuli surrounding the galaxy, and (b) four quadrants in different regimes of the halo (East, North, West, and South). We used unsmoothed, point-source-masked PN images in the 0.3--1.2\,keV band and determined the net count rate for circular segments of width 0\farcm5 between radii of $r=1\arcmin$ and $r=8\arcmin$. For each segment, we summed the counts of all unmasked pixels and determined the mean exposure from the PN exposure map. All count rates were divided by the area of the corresponding segment, taking into account masked point sources and chip gaps. The emission from the $D_{25}$ ellipse of the galactic disk was also excluded. Figure~\ref{xmm_img} (right panel) shows the EPIC image in the 0.3--1.2\,keV band, illustrating the circular segments and quadrants used for the surface-brightness profiles, as well as the distribution of the diffuse emission in these regions.

The background annulus was evaluated at radii of $7\farcm8$ to $9\farcm4$ from the center of NGC~5746. To account for spatial background variations on arcminute scales, we divided the background annulus into forty approximately equal-area patches (${\sim}1.8\,\mathrm{arcmin}^2$) and measured the surface brightness in each patch using the same approach as for the source regions. The scatter of these measurements includes both statistical uncertainties and intrinsic spatial variations of the background. We therefore estimated the background variance term by subtracting the expected Poisson contribution from the total scatter. The background variance term was then added to the statistical uncertainties of the source regions and scaled with extraction area to match the size of each radial bin.

The surface brightness profiles shown in Fig.~\ref{brightness_profiles} represent the background-subtracted count rate per segment, including the $1\sigma$ uncertainty derived from both the statistical errors and the spatial variation of the background count rate. We interpret the halo emission seen in the surface brightness profiles as a significant detection when the count rate exceeds the background at the $1\sigma$ level in two consecutive, radial bins. We quantified the cumulative significance of the extended emission by computing an integrated signal-to-noise ratio (see Eq.~\ref{signif}) for a set of radial bins $i$ (inner-most to outer-most bin with a positive excess):
\begin{equation}\label{signif}
(S/N)_{\rm int}=\frac{\sum_{i} S_i\,A_i}{\sqrt{\sum_{i}(\sigma_i\,A_i)^2}}
\end{equation}
where $S_i$ is the background-subtracted surface brightness in bin $i$, $\sigma_i$ is its $1\sigma$ uncertainty, and $A_i$ is the extraction area.

\subsection{Spectra}\label{sect_spectra}

The surface brightness profiles show that the diffuse X-ray emission in NGC~5746's halo extends to at least ${\sim}3.5\,\arcmin$ from the galactic center. We extracted the spectrum of a $r=5'$ circle centered on the galactic core (excluding the $D_{25}$ ellipse of the galaxy) to investigate the spectral properties of the halo gas.

Furthermore, we identified two regions that are particularly bright in the 0.7--1.2\,keV band, via visual inspection of the X-ray image (Fig.~\ref{xmm_img}). We explored the possibility that this X-ray emission is produced by hot plasma that has been expelled from the disk via stellar feedback. We extracted the spectra for two identical ellipses (semimajor axis $a = 2\arcmin$, semiminor axis $b=1.5\arcmin$) and compared the spectral properties to those of the halo gas.

The EPIC spectra of the disk, the halo, the two regions of enhanced emission (East and West), and the background region were extracted according to the procedure described in Sect.~\ref{chap2}. We first describe our background model and then discuss our analysis of the X-ray spectra.
  
\subsubsection{Background model}\label{sect_bkg}

There are several contributors to background noise in the sensitivity range of \textit{XMM-Newton}. The astrophysical X-ray background is particularly strong at soft energies below ${\sim}1\,\mathrm{keV}$, including a highly variable component produced by charge exchange processes with solar wind particles. The particle background includes contributions from residual soft proton flares and dominates at higher energies. Moreover, we need to account for the instrumental background that is produced by particle interactions with the surroundings of the detectors and the detector material. 

The astrophysical background includes an unabsorbed thermal component (\textsc{apec} model in XSPEC; \citealt{apec}) with a plasma temperature of $kT {\sim} 0.1\,\mathrm{keV}$, accounting for X-ray emission from the local hot bubble and the heliosphere. The temperature of this component is fixed, as the low sensitivity of the EPIC cameras in this energy range does not allow for a confident fit. Secondly, the model requires at least one absorbed thermal component to model the emission from the Galactic halo. The cool component of the Galactic halo typically exhibits temperatures of ${\sim}0.2\,\mathrm{keV}$, while the hot component has a temperature of ${\sim}0.7\,\mathrm{keV}$ \citep{henley2013, bluem2022}.
Lastly, the unresolved emission from extragalactic sources such as active galactic nuclei (AGN), galaxies, and galaxy clusters was modeled by an absorbed power-law with a fixed power-law index of $\Gamma = 1.46$. We used the Tuebingen-Boulder ISM absorption model (\textsc{tbabs} model in XSPEC) with the corresponding abundances from \citet{wilms} to account for Galactic absorption. The hydrogen column density was fixed at $N_\mathrm{H} = 3.3\cdot10^{20}\,\mathrm{cm^{-2}}$ (from \citealt{HI4PI} for the direction of NGC~5746).

Another component that affects most observations is SWCX \citep{swcx}. It is produced by charge exchange processes between ionized solar wind particles and atmospheric hydrogen near the Earth, or ISM particles passing through the solar system. Due to variations in solar wind density and speed, this component is time-variable and can differ between the four observations. We modeled the SWCX contribution with eight Gaussians at line energies of common SWCX lines (according to the ESAS cookbook): a \ion{C}{vi} line at $0.37\,\mathrm{keV}$, \ion{C}{vi} at $0.46\,\mathrm{keV}$, \ion{O}{vii} at $0.57\,\mathrm{keV}$, \ion{O}{viii} at $0.65\,\mathrm{keV}$, \ion{O}{viii} at $0.81\,\mathrm{keV}$, \ion{Ne}{ix} at $0.92\,\mathrm{keV}$, \ion{Ne}{ix} at $1.02\,\mathrm{keV}$ and \ion{Mg}{xi} at $1.35\,\mathrm{keV}$.

The particle background consists of high-energy CRs interacting with the detector material, and solar protons that were diverted into the telescope by its X-ray mirrors. The SP component is highly time-variable and strongest during solar flares. Time intervals affected by SP flares were therefore filtered from the spectra (see Sect.~\ref{sect_processing}). The residual SP emission and CR background is relatively stable and was fitted by an absorbed, broken power-law model.

In addition, high-energy cosmic rays produce fluorescence emission in the telescope through interactions with the detector material and the surrounding structures. The two prominent lines captured by the MOS detectors are the $1.486\,\mathrm{keV}$ Al-K$\alpha$ and $1.739\,\mathrm{keV}$ Si-K$\alpha$ lines. The PN detector captures the Al-K$\alpha$ line, as well as multiple strong lines between 7.2 and $9.2\,\mathrm{keV}$. In addition, we detect a Ti K$\alpha$ line at $4.51\,\mathrm{keV}$ and a Cr K$\alpha$ line at $5.41\,\mathrm{keV}$. The Al and Si lines were modeled by Gaussians.

\begin{figure*}[htbp!]
    \centering
    \includegraphics[width=0.49\linewidth]{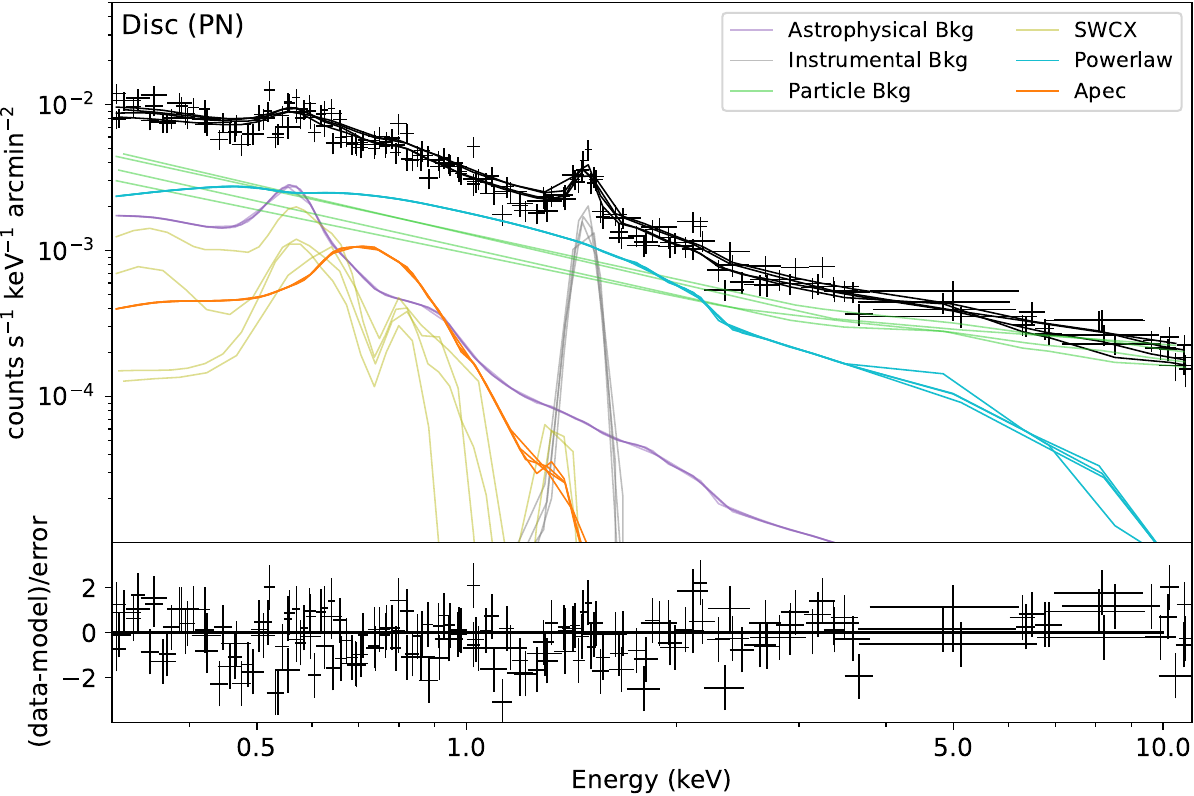}
    \includegraphics[width=0.49\linewidth]{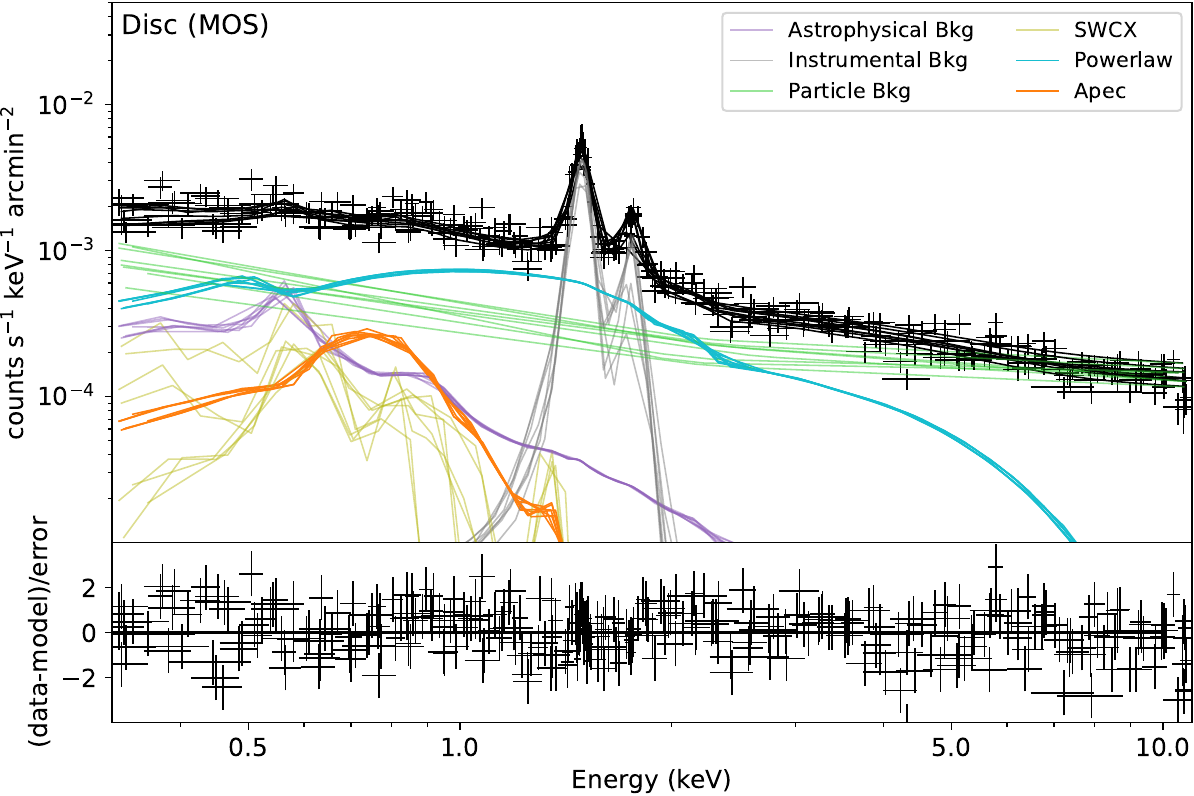}
    \includegraphics[width=0.49\linewidth]{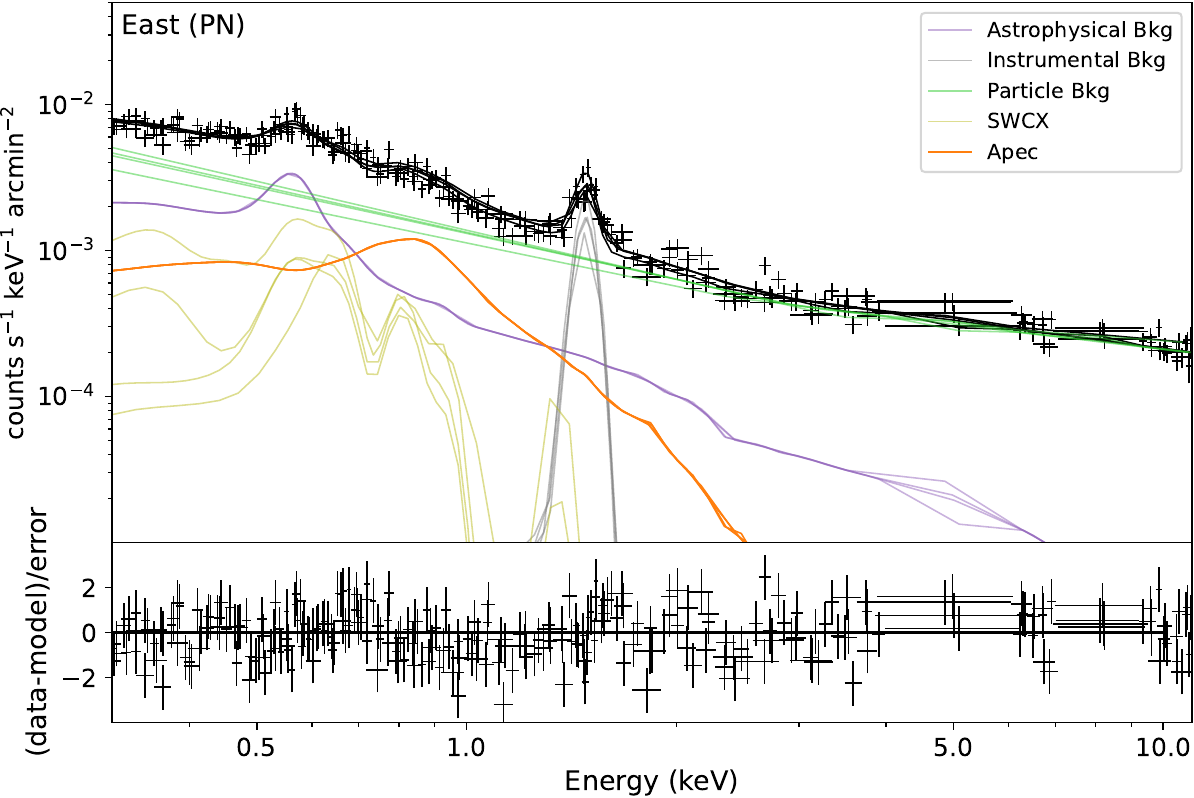}
    \includegraphics[width=0.49\linewidth]{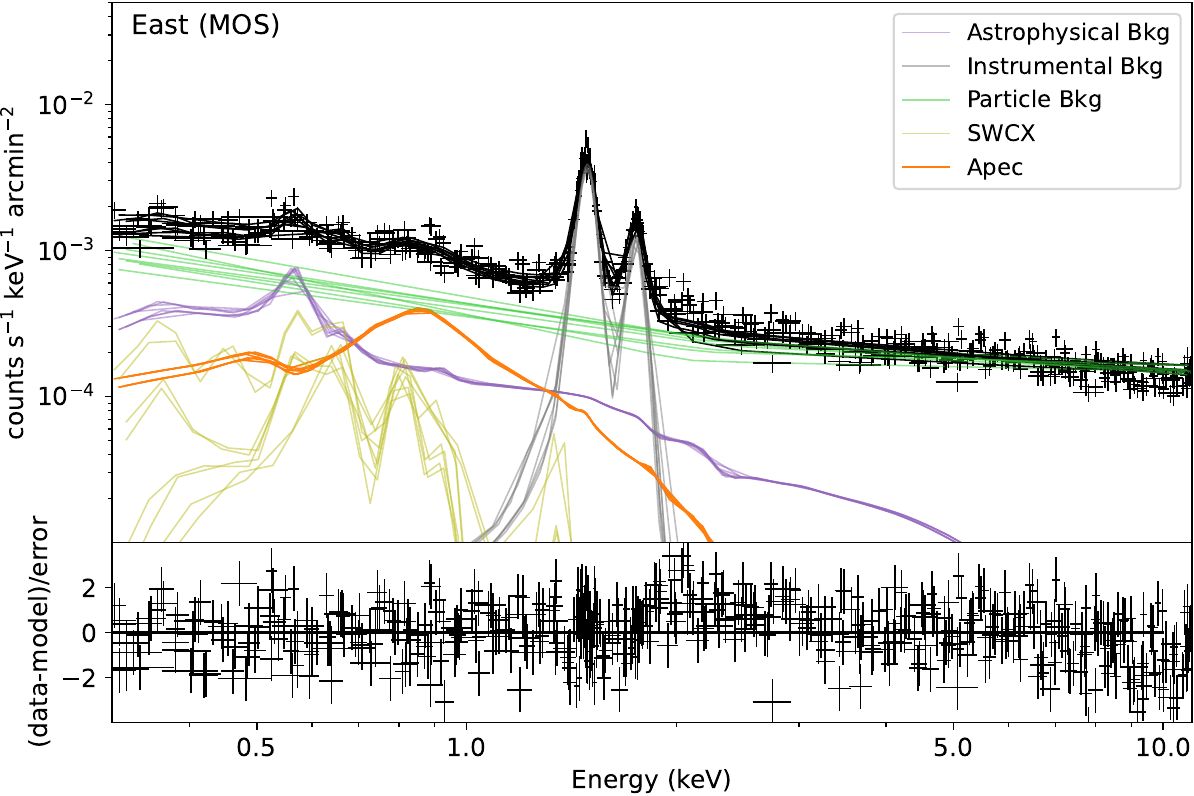}
    \includegraphics[width=0.49\linewidth]{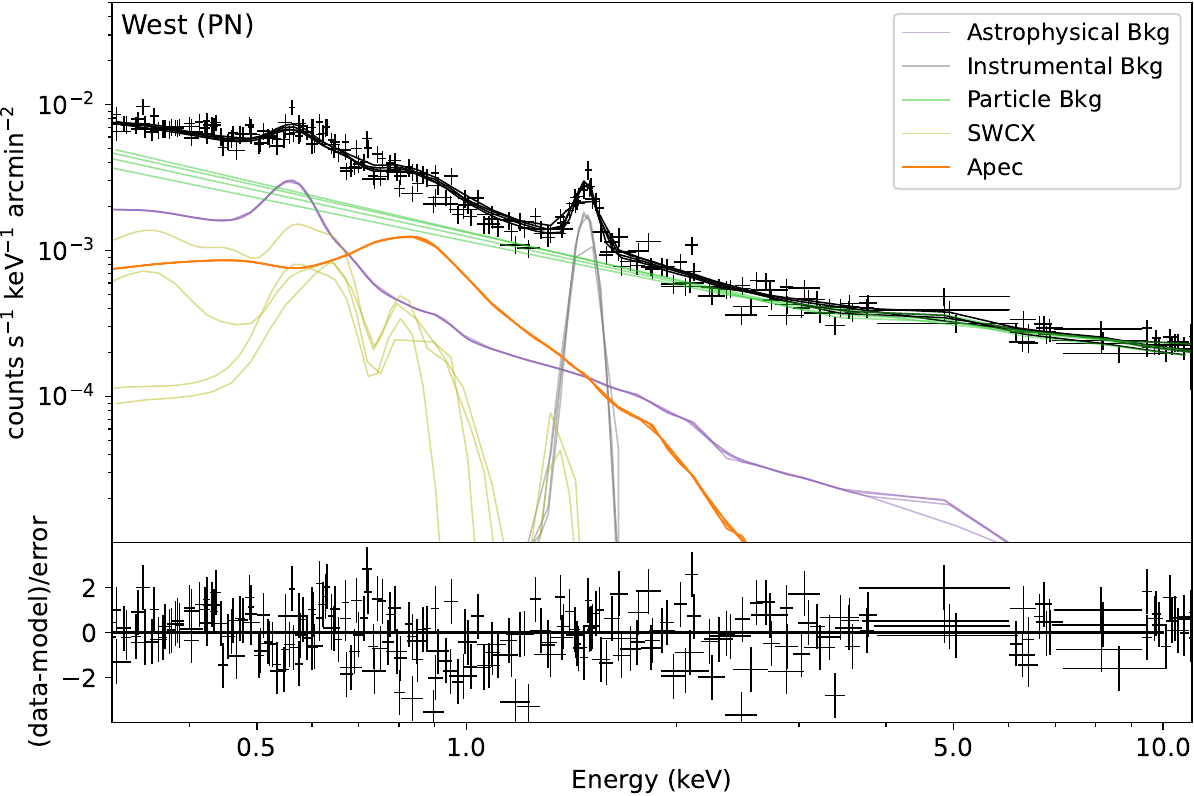}
    \includegraphics[width=0.49\linewidth]{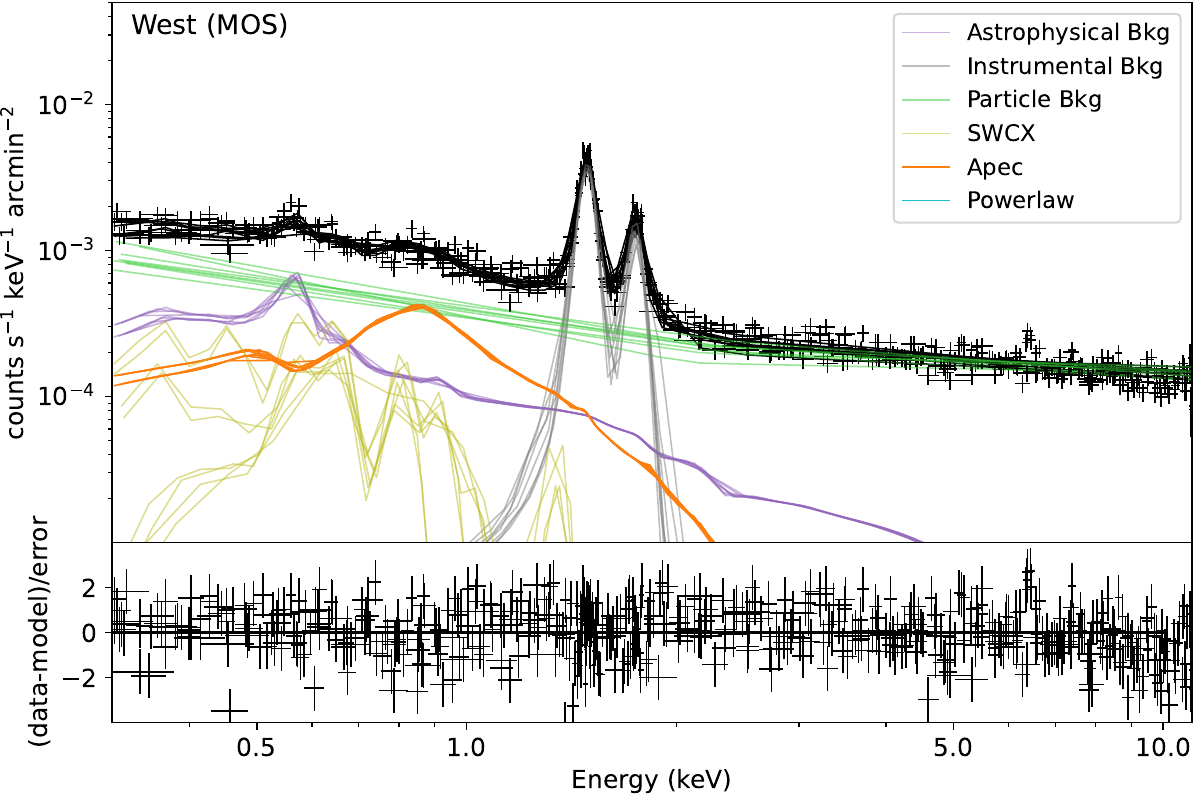}
    \includegraphics[width=0.49\linewidth]{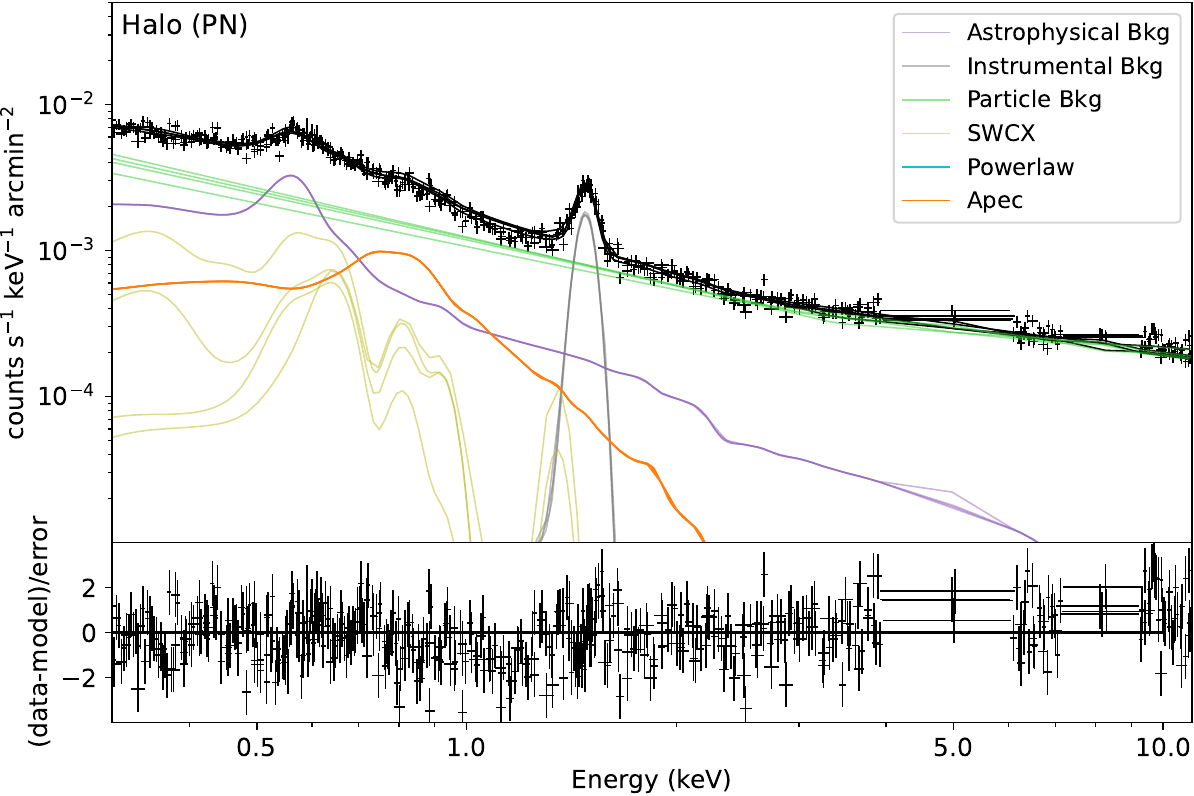}
    \includegraphics[width=0.49\linewidth]{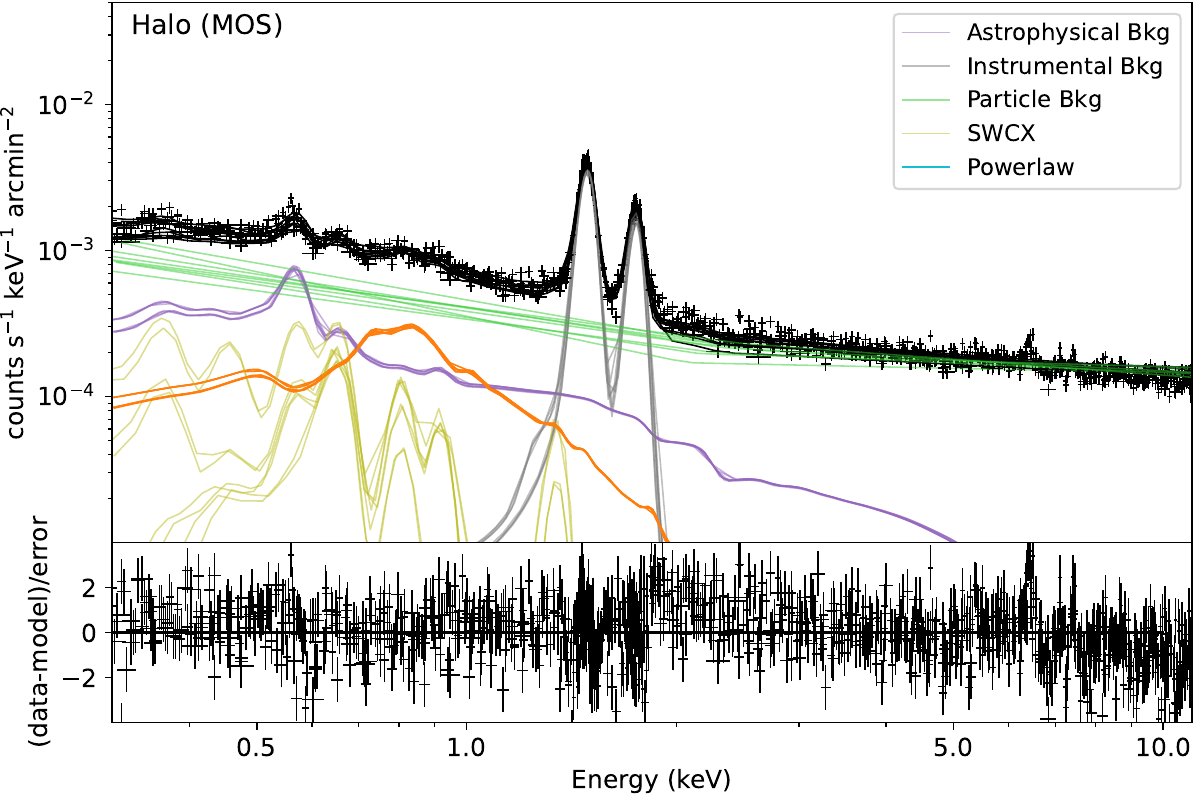}
    \caption{Spectra of PN and MOS1+MOS2 for the disk, the halo out to $r = 5\,\arcmin$, and the two regions of enhanced emission. The data in the plots were binned for better visibility. The colored lines represent the best fits for the different model components.}\label{spectra}
\end{figure*}

    \begin{table*}
        \centering 
        \caption{Results of the spectral modeling.}\label{fit_results}
\begin{tabular}{cccccc}
  \hline\hline             
    Parameter              & Model/Unit                                      & Disk                            & Halo                            & East                            & West \\\hline
    cstat/dof              & \textsc{apec} only                                           &                                 & 1.12                            & 1.10                            & 1.09                            \\\vspace{2pt}
    $kT$                   & $\mathrm{keV}$                            &                                 & $0.62^{+0.04}_{-0.08}$          & $0.75^{+0.07}_{-0.08}$          & $0.73^{+0.07}_{-0.11}$          \vspace{2pt}\\
    $Z/Z_{\odot}$          &                                           &                                 & $0.1$                           & $0.1$                           & $0.1$                           \vspace{2pt}\\
    $\mathrm{norm_{apec}}$ & $\mathrm{cm^{-5}}$                        &                                 & $2.5^{+0.3}_{-0.2}\cdot10^{-6}$ & $3.0^{+0.3}_{-0.3}\cdot10^{-6}$ & $3.4^{+0.4}_{-0.3}\cdot10^{-6}$ \vspace{10pt}\\
    cstat/dof              &  \textsc{apec+powerlaw}                                         & 1.00                            & 1.11                            & 1.10                            & 1.09                            \\\vspace{2pt}
    $kT$                   & $\mathrm{keV}$                            & $0.30^{+0.34}_{-0.06}$          & $0.29^{+0.05}_{-0.03}$          & $0.39^{+0.17}_{-0.09}$          & $0.32^{+0.05}_{-0.05}$          \vspace{2pt}\\
    $Z/Z_{\odot}$          &                                           & $1.0$                           & $0.1$                           & $0.1$                           & $0.1$                           \vspace{2pt}\\
    $\mathrm{norm_{apec}}$ & $\mathrm{cm^{-5}}$                        & $5.9^{+2.7}_{-3.3}\cdot10^{-7}$ & $2.6^{+0.5}_{-0.5}\cdot10^{-6}$ & $1.7^{+0.7}_{-0.6}\cdot10^{-6}$ & $3.6^{+0.9}_{-0.7}\cdot10^{-6}$ \vspace{2pt}\\
    $\Gamma$               &                                           & $1.60^{+0.11}_{-0.09}$          & $1.7$                           & $1.7$                           & $1.7$                           \vspace{2pt}\\
    $\mathrm{norm_{PL}}$   & $\mathrm{cts\,keV^{-1}\,cm^{-2}\,s^{-1}}$ & $2.3^{+0.2}_{-0.2}\cdot10^{-6}$ & $5.2^{+0.5}_{-0.5}\cdot10^{-7}$ & $7.0^{+1.0}_{-0.9}\cdot10^{-7}$ & $5.7^{+0.7}_{-0.9}\cdot10^{-7}$ \\\hline\hline
\end{tabular}
\tablefoot{Fit results for the disk, the halo out to $r = 5\,\arcmin$, and the two regions of enhanced emission. The top section of the table displays the results for the \textsc{apec} only model; the lower section displays the results for the \textsc{apec+powerlaw} model. The metal abundance $Z$ and the power-law indices of the halo regions were fixed to the presented values.}
    \end{table*}

\subsubsection{Source model}\label{src_model}

According to  \textit{Chandra}, \textit{XMM-Newton}, or \textit{Suzaku} studies (e.g., \citealt{intr7,intr8}, and references therein) of diffuse X-ray emission in nearby galaxies (e.g., the Magellanic Clouds or M31) and our own Galaxy, the diffuse emission in most galaxies is consistent with thermal plasma emission at temperatures of $kT {\sim}0.2\,\mathrm{keV}$. This emission component most likely originates from the hot ISM in equilibrium and unresolved stellar sources. In addition, many galaxies also display hotter areas such as \ion{H}{ii} regions, SNRs, and superbubbles. The contribution of these regions can be modeled by a second thermal plasma component with a temperature of $kT > 0.5\,\mathrm{keV}$. 

In addition, we expect a nonnegligible contribution by unresolved X-ray binaries (XRBs) to the spectrum of the galactic disk. \citet{mineo2} studied the emission from the hot ISM and the populations of high-mass X-ray binaries (HMXBs) in a sample of 29 nearby star-forming galaxies using \textit{Chandra}, \textit{Spitzer}, \textit{GALEX}, and 2MASS data. They found that, on average, ${\sim}30-40\%$ of the galaxies' apparent luminosity from diffuse emission in the $0.5-2\,\mathrm{keV}$ energy range is produced by faint compact sources, mainly XRBs.

We modeled the spectrum of the galactic disk by a combination of one thermal plasma component and a power-law component, as the photon statistics are insufficient to distinguish between a hot and a cold thermal component.
We attempted to fit the spectrum with the nonequilibrium ionization \textsc{nei} model, as the contribution of SNRs to the diffuse X-ray emission in star-forming galaxies can be quite significant. However, we found that the fit results are mostly independent from the ionization timescale $\tau$ and the data are better fit by the collisional ionization equilibrium \textsc{apec} model. The source components were multiplied by a (second) \textsc{tbabs} absorption model to account for intrinsic absorption in NGC~5746. The combined model for the astrophysical background and the source emission looks as follows: \textsc{constant(apec + tbabs(apec + apec + powerlaw + tbabs(apec + powerlaw)))}, where the multiplicative factor scales each observation by the effective area covered in the sky. Although we do not expect many XRBs to be present outside the disk, the halo spectra might contain residual, nonthermal emission from point sources that were not fully masked or resolved. We therefore tested for a possible nonthermal contribution by performing fits both with and without an additional power-law component.

\subsubsection{Spectral analysis}

We simultaneously fitted the MOS and PN data of all four observations in the 0.3--11.0\,keV energy range. For the PN data, we excluded the 4.0--6.0\,keV and 7.2--9.2\,keV energy ranges due to the presence of instrumental lines. We did not bin the data and used cstat statistics. For each camera and observation, we calculated the point source- and chip-gap-masked extraction area from the \textsc{backscal} header of the spectral file. The value for the area in $\text{arcmin}^2$ was inserted for the constant model factor to cross-calibrate between observations.

{We first fitted the background model to the background region. The astrophysical background parameters were linked between detectors and observations, while the particle and instrumental background components were fitted independently for each detector and observation, since they are detector-dependent and vary between exposures. The SWCX line normalizations were linked between the different EPIC detectors, but not between observations, as they are time-variable. 

In a second step, we performed a joint fit of the source and the background spectra. The background model was applied to the source spectra with the normalizations scaled according to the ratio of the extraction areas. The shape of the background model was not kept fully fixed in the source fit; while the spectral shapes of the background components were fixed to the best-fit values from the background fit, we allowed the normalizations of all background components to vary during the joint fit. This approach propagates the most relevant uncertainties in the background level to the uncertainties of the source parameters.

We fitted the source model to the disk spectrum. The element abundance is poorly constrained by the disk spectrum and was therefore fixed to the interstellar value of \citet{wilms}. The analysis of the halo regions (Halo, East, and West) is more challenging due to the low surface brightness of the diffuse emission compared to the background components. The spectra are dominated by a strong particle background (see Fig.~\ref{spectra}), and the photon statistics are insufficient to resolve emission line features, leading to a continuum-dominated spectral shape. We therefore adopted a single-temperature \textsc{apec} model with the abundance fixed to $Z = 0.1\,\mathrm{Z_{\odot}}$ (see, e.g., \citealt{hodges2013}). To test the sensitivity of the plasma temperature to residual nonthermal emission (e.g., unresolved point sources), we additionally fitted an alternative model including a power-law component. For the disk region, the power-law index was fitted freely, while for the halo regions, it was fixed to $\Gamma = 1.7$. This value lies within the range typically observed for unresolved AGN and compact X-ray source populations ($\Gamma {\sim} 1.4\mbox{--}2.0$; e.g., \citealt{deluca2004,mineo1}).

\subsection{Fluxes}\label{sect_fluxes}

Using the flux command in XSPEC, we determined the X-ray fluxes for multiple energy bands: soft (0.3--0.7\,keV), medium (0.7--1.2\,keV), and hard (1.2--4.0\,keV). We also calculated the flux in 0.3--2.0\,keV to compare our results to the literature.

Each observation covers a slightly different area in the sky, since (a) the cameras have different chip gaps, and (b) the source masks look slightly different for each observation due to the varying exposure times. We therefore calculated the area-weighted mean flux $F$ for each region and energy band. To determine the flux of a specific model component, we simply set the normalizations of all other model components to zero. To determine the absorption-corrected flux, we set the absorbing hydrogen column density $N_\mathrm{H}$ to zero. The flux errors were then propagated from the normalization error and the statistical error introduced by the calculation of the area-weighted mean flux (see Eq. \ref{eq_flux}).

\begin{equation}\label{eq_flux}
F = \frac{\sum_\mathrm{i} A_\mathrm{i}\,F_\mathrm{i}}{\sum_\mathrm{i} A_\mathrm{i}}\;\;\;\;\;\;\;\;\;\;          \Delta F = \sqrt{\left(\frac{\Delta_\mathrm{norm}}{\mathrm{norm}}F\right)^2+\frac{\sum_\mathrm{i} A_\mathrm{i}(F-F_\mathrm{i})^2}{\sum_\mathrm{i}A_\mathrm{i}}}.
\end{equation}

The term $\Delta \text{norm}$ represents the error of the model normalization, $F_\mathrm{i}$ is the flux, and $A_\mathrm{i}$ is the effective area covered by one observation.

\section{Results}\label{chap4}

In this section, we report the measured extent, morphology, temperatures, and luminosities of the diffuse emission. Interpretations of the origin of the halo are discussed in Sect.~\ref{chap5}.

The spectra are shown in Fig.~\ref{spectra}. The fit results for the disk, the halo, and the two regions of enhanced emission (East and West) are presented in Table~\ref{fit_results}. The X-ray fluxes, calculated for different energy bands, are listed in Table~\ref{tab_fluxes}.

    \begin{table*}
        \centering 
        \caption{X-ray fluxes.}\label{tab_fluxes}
        \begin{tabular}{cccccc} \hline\hline        
          Region   & Model         & $F_\mathrm{0.3-2.0\,keV}$        & $F_\mathrm{0.3-0.7\,keV}$        & $F_\mathrm{0.7-1.2\,keV}$        &  $F_\mathrm{1.2-4\,keV}$\\\hline

          Disk & \textsc{apec}     & $4.6^{+2.1}_{-2.6}\cdot10^{-15}$ & $2.2^{+1.1}_{-1.3}\cdot10^{-15}$ & $2.2^{+1.0}_{-1.3}\cdot10^{-15}$ & $2.4^{+1.1}_{-1.4}\cdot10^{-16}$  \vspace{2pt}\\
               & \textsc{powerlaw} & $3.2^{+0.3}_{-0.3}\cdot10^{-14}$ & $1.1^{+0.1}_{-0.1}\cdot10^{-14}$ & $9.6^{+0.7}_{-0.7}\cdot10^{-15}$ & $3.1^{+0.3}_{-0.3}\cdot10^{-14}$  \vspace{2pt}\\
               & Total             & $3.7^{+0.4}_{-0.4}\cdot10^{-14}$ & $1.4^{+0.2}_{-0.2}\cdot10^{-14}$ & $1.2^{+0.2}_{-0.2}\cdot10^{-14}$ & $3.1^{+0.3}_{-0.3}\cdot10^{-14}$  \vspace{10pt}\\

          Halo & \textsc{apec}     & $8.7^{+1.7}_{-1.7}\cdot10^{-14}$ & $3.6^{+0.7}_{-0.7}\cdot10^{-14}$ & $4.0^{+0.8}_{-0.8}\cdot10^{-14}$ & $1.3^{+0.3}_{-0.3}\cdot10^{-14}$  \vspace{2pt}\\
          
          East & \textsc{apec}     & $2.5^{+1.1}_{-0.9}\cdot10^{-14}$ & $9.8^{+4.1}_{-3.5}\cdot10^{-15}$ & $1.1^{+0.5}_{-0.4}\cdot10^{-14}$ & $5.2^{+0.3}_{-0.2}\cdot10^{-15}$  \vspace{2pt}\\

          West & \textsc{apec}     & $2.7^{+0.7}_{-0.6}\cdot10^{-14}$ & $1.1^{+2.7}_{-2.1}\cdot10^{-14}$ & $1.2^{+0.4}_{-0.3}\cdot10^{-14}$ & $5.6^{+1.4}_{-1.1}\cdot10^{-15}$  \vspace{2pt}\\\hline
        \end{tabular}
\tablefoot{Absorption-corrected X-ray fluxes in $\mathrm{erg\,s^{-1}\,cm^{-2}}$ for the disk, the halo out to $r = 5\,\arcmin$, and the two regions of enhanced emission (East and West). Point sources have been masked out.}
    \end{table*}

\subsection{The disk}

For the galactic disk, we measure a plasma temperature of $kT = 0.30^{+0.34}_{-0.06}\,\mathrm{keV}$. The emission in the disk is dominated by the nonthermal component, while only ${\sim}12\,\%$ of the absorption-corrected flux in 0.3--2.0\,keV is attributed to the thermal component. The 0.3--2.0\,keV flux of $(3.7 \pm 0.4) \cdot 10^{-14}\,\mathrm{erg\,s^{-1}\,cm^{-2}}$ corresponds to a luminosity of $(3.7 \pm 0.4) \cdot 10^{39}\,\mathrm{erg\,s^{-1}}$ at a distance of 29\,Mpc.

Previous studies of star-forming disks have shown that diffuse emission can often be described by at least two thermal components (see Sect.~\ref{src_model}). Typically, these consist of a cooler ($kT {\sim}0.2\,\mathrm{keV}$) and a hotter ($kT > 0.5\,\mathrm{keV}$) component. Unfortunately, the quality of the present EPIC data does not allow us to resolve multiple thermal components in addition to the nonthermal emission. We therefore modeled the diffuse emission with a single-temperature plasma. The resulting plasma temperature has a broad uncertainty range, which arises due to the limited photon statistics, as well as the degeneracy with the nonthermal component.

\subsection{The halo}

The surface-brightness profiles show significant, diffuse X-ray emission extending beyond the galactic disk into the halo of NGC 5746. In the azimuthally averaged profile (left panel in Fig.~\ref{brightness_profiles}), the halo emission is detected out to a projected radius of ${\sim}3.5\arcmin$, corresponding to ${\sim}29\,\mathrm{kpc}$. While one bin at $r{\sim}1.5\mbox{--}2\arcmin$ is consistent with the background within the uncertainties, the integrated signal remains significant at the ${\sim}3\sigma$ level.

The quadrant profiles (Fig.~\ref{brightness_profiles}, right) show pronounced azimuthal variations. In the North and South, no statistically significant excess is detected. In the eastern quadrant, a local enhancement is seen in a single radial bin at $r{\sim}2.5\mbox{--}3\arcmin$, with a significance of ${\sim}1.6\sigma$. The western quadrant shows a broader excess between $r{\sim}1.5\mbox{--}3.5\arcmin$, corresponding to four consecutive bins and an integrated significance of $\sim3.7\sigma$. The western quadrant exhibits another excess at $r{\sim}6.5\mbox{--}7.5\arcmin$, which is likely associated with residual emission from background sources. According to the SIMBAD astronomical database\footnote{\url{https://simbad.u-strasbg.fr/simbad/sim-fbasic}}, multiple galaxies, such as SDSS J144427.31+015658.8, and a foreground star are located at this position. This excess may therefore arise from unresolved point sources or residual emission from the wings of the point spread function.

We extracted the halo spectrum in a circular region with radius $r = 5\arcmin$, excluding the $D_{25}$ optical extent of the disk. This radius was chosen to enclose the full extent of the halo emission in the surface-brightness profiles, while excluding regions where the signal becomes dominated by background emission. Still, we find that the halo spectrum has a low surface brightness and is dominated by background emission, resulting in a continuum-dominated shape (see bottom row in Fig.~\ref{spectra}). The photon statistics are insufficient to disentangle multiple thermal components or to resolve individual emission lines. When left free, the metal abundance is not constrained by the fit and converges toward very low values ($Z {\sim} 0.02$\mbox{--}$0.03,\mathrm{Z_{\odot}}$). We therefore modeled the halo emission with a single-temperature thermal plasma model and fixed the metal abundance to $Z = 0.1\,\mathrm{Z_{\odot}}$.

We find a plasma temperature in the halo of $kT = 0.62^{+0.04}_{-0.08}\,\mathrm{keV}$ and a luminosity of $8.8^{+1.7}_{-1.7}\cdot10^{39}\,\mathrm{erg\,s^{-1}}$ in 0.3--2.0\,keV. To assess the sensitivity of the plasma temperature to the chosen model, we additionally fitted an alternative model including a power-law component, i.e., to account for residual emission from masked or unresolved point sources. When the photon index of the power-law component is allowed to vary freely, it becomes unconstrained and pegs at the hard lower limit ($\Gamma = 1.0$). The photon index was therefore fixed to $\Gamma = 1.7$. The addition of a power-law component leads to significantly lower best-fit plasma temperatures; however, it does not result in a statistically significant improvement of the fit quality and produces strong degeneracies between the thermal and nonthermal components. We therefore adopted the single-temperature thermal model as our reference description of the halo emission. The differences between the models are treated as an estimate of the systematic uncertainty introduced by the spectral modeling.

\subsection{The regions of enhanced emission}\label{sect_bubbles}

The surface brightness profiles and X-ray images reveal two regions of enhanced extraplanar emission to the East and West of the galactic disk (cyan ellipses in the left panel of Fig.~\ref{xmm_img}). The absorption-corrected 0.3--2.0\,keV luminosities of the diffuse thermal emission are $2.5^{+1.1}_{-1.0}\cdot10^{39}\,\mathrm{erg\,s^{-1}}$ for the eastern region and $2.7^{+0.7}_{-0.6}\cdot10^{39}\,\mathrm{erg\,s^{-1}}$ for the western region (see Table~\ref{tab_fluxes}).

In the right panel of Fig.~\ref{xmm_img}, we show a 0.3--1.2\,keV image of NGC 5746. It uses a narrower percentile stretch than the full 0.3--5\,keV image, which suppresses soft background variations and highlights the spatial structure of the extraplanar emission. In this representation, the extraplanar emission appears rather filamentary than smooth, extending away from the galactic plane. In the full-band image, the eastern region shows multiple blue spots (hard band; 1.2--5.0\,keV) that could either represent residual emission from point sources or emission from a strongly heated gas. 

Using the same spectral modeling approach as for the halo, we measure plasma temperatures of $0.75^{+0.07}_{-0.08}\,\mathrm{keV}$ in the eastern region and $0.73^{+0.07}_{-0.11}\,\mathrm{keV}$ in the western region. These temperatures are slightly higher than the best-fit values obtained for the averaged halo. However, due to the limited photon statistics, the uncertainties are substantial. As in the case of the halo, alternative models including a power-law component lead to lower best-fit plasma temperatures, indicating that the temperatures are sensitive to the chosen model.

\section{Discussion}\label{chap5}

\subsection{Comparison to former studies}

As we mentioned in the introduction, NGC~5746 has been subject to multiple X-ray studies in the past. \citet{pedersen2006} were the first to present the discovery of extended X-ray emission around the galaxy. Using a $36.8\,\mathrm{ks}$ \textit{Chandra} ACIS observation, these authors detected soft, diffuse gas extending more than $20\,\mathrm{kpc}$ from the disk. In a more recent publication \citep{rasmussen2009}, however, the authors retracted these findings. \citeauthor{rasmussen2009} argued that the detected signal was introduced by the mishandling of the spatial variations in the quantum efficiencies of the ACIS CCDs. Applying newer calibration data, they found no significant evidence of a diffuse X-ray halo in NGC~5746.

Our analysis of the EPIC data, based on both imaging and surface brightness profiles, presents evidence of diffuse X-ray emission extending out to projected radii of approximately 30\,kpc. The discrepancy with the \textit{Chandra} results can be explained by instrumental differences. The \textit{Chandra} ACIS has a smaller FOV and a smaller effective area (${\sim}290\,\text{arcmin}^2$ and ${\sim}400\,\mathrm{cm}^2$ at $1\,\mathrm{keV}$; see Chandra Proposer's Observatory Guide\footnote{\url{https://cxc.harvard.edu/proposer/POG/html/chap6.html}}) compared to \textit{XMM-Newton's} EPIC (${\sim}630\,\text{arcmin}^2$ and ${\sim}1800\,\mathrm{cm}^2$ at $1\,\mathrm{keV}$; \citealt{pn, mos}). Hence, while \textit{Chandra} is very sensitive to faint point sources, the detection of faint, diffuse emission is difficult with \textit{Chandra}; \textit{XMM-Newton}, on the other hand, is far more sensitive and efficient in the soft X-ray band for picking up a diffuse signal. In addition, the effective exposure of the EPIC observations is significantly longer, with $248\,\mathrm{ks}$ for the MOS and $205\,\mathrm{ks}$ for the PN detector. The nondetection reported by \citet{rasmussen2009} can therefore be attributed to the sensitivity limits of the available \textit{Chandra} observations.

Diffuse X-ray halos are commonly observed in starburst galaxies, where the high rate of SN explosions and stellar winds can drive powerful galactic outflows that are bright in X-rays \citep[e.g.,][]{strickland2004a}. In contrast, detections of extended X-ray halos around normal, star-forming galaxies remain relatively scarce (e.g., NGC~891; \citealt{bregman1994}). The luminosity of extraplanar X-ray emission has been found to correlate with the SFR of the disk \citep{tüllmann2006b}, which underlines the importance of stellar feedback for the formation of X-ray halos.

Based on a statistical analysis of late-type galaxies, \citet{tüllmann2006b} identified an approximate threshold of $\text{SFR} \gtrsim 1.0\,\mathrm{M_{\odot}\,yr^{-1}}$ for the detection of multiphase halos. However, this relation exhibits a large scatter and does not account for the halo mass of the galaxy. Recent studies have shown that massive spiral galaxies can host hot, X-ray emitting halos even at lower SFRs (e.g., \citealt{li2017}).

\subsection{The star formation rate in the disk}

Among the other possible mechanisms, stellar feedback often plays an important role in producing extraplanar X-ray emission. It is therefore important to assess whether the present star formation activity in NGC~5746 could possibly contribute to the observed X-ray halo. The HMXBs are a direct tracer of recent star formation, as their luminosity is linked to the lifetime of massive OB stars. The integrated HMXB luminosity $L_\mathrm{HMXB}$ in star-forming galaxies is therefore expected to scale with the SFR. We adopt the empirical relation following \citet{mineo1} to estimate the SFR (see Eq.~\ref{lx_sfr}).
\begin{equation}\label{lx_sfr}
    L_\mathrm{HMXB}\,\,(\mathrm{erg\,s^{-1}}) \sim 2.61\cdot10^{39}\times \mathrm{SFR}\,\,(\mathrm{M_{\odot}\,yr^{-1}})
\end{equation}
We fitted the X-ray spectrum of the disk without masking point sources, and determined the 0.5--8\,keV flux of the nonthermal (power-law) component similar to the procedure described in Sect.~\ref{sect_fluxes}. Assuming a distance of 29\,Mpc, we obtain a nonthermal X-ray luminosity of $L_\mathrm{0.5-8\,keV} = (2.53^{+0.18}_{-0.17}) \cdot 10^{40}\,\mathrm{erg\,s^{-1}}$. Distance measurements for NGC 5746 vary significantly in the literature. Adopting $d = (29 \pm 5)\,\mathrm{Mpc}$ \citep{hecate} introduces an additional systematic uncertainty of ${\sim}34\,\%$ in luminosity (and thus SFR). Studies conducted prior to 2010 list distances of $d < 29\,\mathrm{Mpc}$ (e.g., \citealt{tully2007}), while more recent studies consistently determine distances $d > 29\,\mathrm{Mpc}$ (e.g., \citealt{tully2016, karachentsev2014}), with the highest estimate reaching $d {\sim}35\,\mathrm{Mpc}$ \citep{sorce2014}. We adopt $d = 29\,\mathrm{Mpc}$ for the following calculation, noting that a larger distance would increase the inferred luminosity and SFR.

Since we expect NGC~5746 to be a quiescent, early-type spiral, its X-ray luminosity is likely dominated by low-mass X-ray binaries (LMXBs). Cataclysmic variables (CVs), and active binaries (ABs) also contribute to the unresolved X-ray emission. We used a stellar mass of $M_{\ast} = (1.2 \pm 0.1) \cdot 10^{11}\,\mathrm{M_{\odot}}$ \citep{jiang2019} to estimate these contributions to the nonthermal luminosity. Using the relation established by \citet{gilfanov2004} for the luminosity of LMXBs with $L > 10^{37}\,\mathrm{erg\,s^{-1}}$ per unit stellar mass, $L_\mathrm{LMXB}/M_{\ast} = (8.0 \pm 0.5)\cdot 10^{28}\,\mathrm{erg\,s^{-1}\,M_{\odot}^{-1}}$, and applying the recommended correction factor of 1.1 to account for LMXBs below the detection threshold, we obtain $L_\mathrm{LMXB} = (1.06 \pm 0.11)\cdot 10^{40}\,\mathrm{erg\,s^{-1}}$. The remaining unresolved emission from CVs and ABs can be estimated using the emissivity measured by \citet{revnivtsev2008}, $L_\mathrm{CV+AB}/M_{\ast} = (1.5 \pm 0.4)\cdot 10^{28}\,\mathrm{erg\,s^{-1}\,M_{\odot}^{-1}}$ in 0.5--7\,keV, which yields $L_\mathrm{CV+AB} = (1.8 \pm 0.5)\cdot 10^{39}\,\mathrm{erg\,s^{-1}}$, assuming that the contribution above 7\,keV is negligible.

Subtracting the estimated contributions of LMXBs, CVs, and ABs from the nonthermal luminosity leaves a residual luminosity of $L_\mathrm{HMXB} = (1.29 \pm 0.21)\cdot 10^{40}\,\mathrm{erg\,s^{-1}}$, which we interpret as the approximate luminosity attributable to HMXBs. Plugging this value into the \citet{mineo1} relation (Eq.~\ref{lx_sfr}) yields a SFR of $4.9^{+0.9}_{-0.8}\,\mathrm{M_{\odot}\,yr^{-1}}$ at the adopted distance of 29\,Mpc. The varying distance estimates listed in the NASA/IPAC Extragalactic Database\footnote{\url{https://ned.ipac.caltech.edu/}} ($17.1\,\mathrm{Mpc} < d < 34.7\,\mathrm{Mpc}$) correspond to $\mathrm{SFR}{\sim}1.7\mbox{--}7.0\,\mathrm{M_{\odot}\,yr^{-1}}$. We emphasize, however, that this estimate is subject to substantial systematic uncertainties. \citet{mineo1} reported a dispersion of $\sigma = 0.43\,\text{dex}$ around their scaling relation (Eq.~\ref{lx_sfr}), and the XRB luminosity of an individual galaxy can deviate from the mean relation due to stochastic sampling of the luminosity function and the variability of bright sources. In addition, our separation of HMXBs from other compact sources relies on stellar-mass based scaling relations, which introduce additional uncertainties. We therefore treated the XRB-based SFR purely as a consistency check on the level of recent star formation. Dust-reprocessed IR emission is generally a more robust tracer of ongoing star formation, as it is less sensitive to the stochastic sampling of bright sources.

Nevertheless, the XRB-based estimate suggests that the recent SFR in NGC 5746 could be higher than the ${\sim}0.8\mbox{--}1.0\,\mathrm{M_{\odot}\,yr^{-1}}$ inferred from IR measurements \citep{jiang2019}. A higher SFR would reduce the  tension with the empirical trend reported by \citet{tüllmann2006b}, who found that multiphase halos are more commonly detected in galaxies with $\text{SFR} \gtrsim 1.0\,\mathrm{M_{\odot}\,yr^{-1}}$. Further multiwavelength constraints are required to robustly determine the present-day SFR.

\subsection{Possible origins of the extraplanar emission}

NGC~5746 is a massive spiral galaxy with an IR-inferred SFR of ${\sim}0.8\mbox{--}1.0\,\mathrm{M_{\odot}\,yr^{-1}}$ \citep{jiang2019}. A possible explanation for the detection of diffuse, extraplanar emission is a quasi-hydrostatic halo that is primarily supported by the galaxy's gravitational potential. Such a halo could have been affected by multiple mechanisms, such as heating associated with stellar feedback from current or past star formation, or the accretion of intergalactic material. Additional contributions from a low-luminosity AGN cannot be excluded.

\subsubsection{Quasi-hydrostatic halo}

A quasi-hydrostatic halo is expected in a massive spiral galaxy such as NGC 5746, whose stellar mass is $M_{\ast}{\sim}(1.1-1.3)\cdot10^{11}\,\mathrm{M_{\odot}}$ \citep{jiang2019}. In such massive systems, the gravitational potential is expected to be deep enough to accumulate and preserve large amounts of hot gas. X-ray observations of similarly massive systems (e.g., UGCA145 in \citealt{li2017}) have revealed extended halos with similar or higher temperatures (${\sim}0.4\mbox{--}1.1\,\mathrm{keV}$), also in galaxies with lower SFRs. Comparable halo temperatures have also been reported for other massive spiral galaxies, such as M104 \citep{li2011} and NGC 2613 \citep{li2006}, as well as in broader samples of massive spirals \citep{li2018}. In this scenario, the measured plasma temperatures are consistent with gas heated to near the virial temperature of the dark matter halo (e.g., \citealt{white1991}).

Quasi-hydrostatic halos typically exhibit a relatively smooth surface-brightness distribution (e.g., \citealt{bogdan2013}). The gas distribution seen in the X-ray image (right panel in Fig.~\ref{xmm_img}) and the surface brightness profiles (right panel in Fig.~\ref{brightness_profiles}) suggests that additional processes may contribute to the observed emission.

\subsubsection{Stellar feedback and outflows}

Stellar feedback is a plausible mechanism for the production of extraplanar X-ray emission in star-forming galaxies. In NGC~5746, the anisotropy and the structure of the detected emission are features that can be observed in galaxies with galactic winds \citep{strickland2004a, strickland2004b, veilleux2005}.

The present-day SFR, inferred from IR indicators ($0.8\mbox{-}1.0\,\mathrm{M_{\odot}\,yr^{-1}}$; \citealt{jiang2019}) is moderate, but the XRB-based estimate ($4.9^{+0.9}_{-0.8}\,\mathrm{M_{\odot}\,yr^{-1}}$) suggests that the recent star formation activity could have been higher, with substantial uncertainties. The cooling mechanisms for hot, X-ray emitting plasma are very inefficient and the cooling timescales can exceed $\tau \gtrsim  10^9\,\mathrm{yrs}$ (e.g., \citealt{bogdan2013, crain2010}). It is therefore plausible that the heated material remains detectable, while the responsible star formation episode already ended. The plasma temperatures inferred from the spectral analysis $kT{\sim}0.6\mbox{--}0.7\,\mathrm{keV}$ are compatible with shock-heated gas generated by outflows (e.g., \citealt{strickland2004a}), but the sensitivity to the adopted model limits their diagnostic power. Stellar feedback is a possible scenario for the observed morphology and spectral properties  of the extraplanar halo.

\subsubsection{Accretion of intergalactic material}

Accretion of gas from the IGM provides another plausible mechanism through which a hot halo may be built up or maintained. In this scenario, infalling material is shock-heated to temperatures near the virial temperature of the dark matter halo and produces diffuse X-ray emission. We know from multiwavelength-studies (e.g., \citealt{tüllmann2006a, strickland2004a}) that X-ray halos in galaxies with active outflows correlate with H$\alpha$ emission from the diffuse ionized gas and with radio continuum emission from high-energy CRs. \citet{rasmussen2009} compared their X-ray data to an optical observation conducted with the $1.54\,\mathrm{m}$ telescope at La Silla, and radio data from the $1.4\,\mathrm{GHz}$ NRAO VLA Sky Survey \citep{condon1998}. They found no evidence of significant H$\alpha$, or radio emission in the halo of NGC~5746. The authors favored the interpretation of intergalactic accretion due to the absence of detectable H$\alpha$ and radio continuum emission in the halo of NGC 5746; the lack of such detections was viewed as evidence against ongoing stellar feedback.

Accretion-driven halos are predicted by theoretical models (e.g., \citealt{benson2000, toft2002}) for massive spiral galaxies. However, the absence of H$\alpha$ and radio emission does not uniquely favor accretion. These tracers are relatively short-lived and limited by the sensitivity of the observation. In normal, star-forming galaxies, radio halos are typically detected out to distances of ${\sim 10\,\mathrm{kpc}}$ (e.g., \citealt{strickland2004a, tüllmann2006a}), while the X-ray halo of NGC 5746 extends out to ${\sim}30\,\mathrm{kpc}$. Moreover, accretion alone would generally be expected to produce an isotropic and smoothly distributed halo (e.g., \citealt{toft2002}), whereas the observed X-ray morphology appears more structured. While accretion from the IGM may contribute to the hot gas reservoir, the data do not allow accretion to be identified as the sole origin of the extraplanar emission.

\subsubsection{AGN feedback}

In principle, it is also possible that the extraplanar gas was heated via AGN feedback. In the case of NGC~5746, however, there is little evidence of AGN activity. \citet{pedersen2006} only detected a very faint X-ray source ($L_\mathrm{X} {\sim} 2\cdot10^{40}\,\mathrm{erg\,s^{-1}}$) in the center of NGC~5746; much too weak for a dormant AGN. The only indication for the presence of a low-luminosity AGN is the classification of NGC~5746 as a low-ionization nuclear emission-line region galaxy by \citet{carillo1999}. Moreover, \citet{strickland2004a}, who analyzed \textit{Chandra} data of a sample of edge-on galaxies with AGN, found that there is no correlation between the presence of an AGN, and the luminosity of the X-ray halo. AGN feedback is therefore unlikely to be the dominant mechanism responsible for the observed extraplanar emission.

\section{Conclusion}\label{chap6}

We have presented a deep \textit{XMM-Newton} study of the massive, edge-on spiral galaxy NGC~5746 based on an effective exposure of ${\sim} 250\,\mathrm{ks}$. The data provide substantially improved sensitivity to faint, diffuse X-ray emission compared to previous observations and enables a detailed study of the spatial extent, morphology, and spectral properties of NGC 5746's hot, circumgalactic gas.

Our imaging analysis and surface brightness profiles present evidence of the detection of extraplanar gas extending out to projected radii of ${\sim}30\,\mathrm{kpc}$. The emission appears to be asymmetric, with the most significant excess detected in the western quadrant between ${\sim}1.5\arcmin$ and ${\sim}3.5\arcmin$, while the eastern side shows only a marginal enhancement and the northern and southern quadrants remain consistent with the background. This detection supports the presence of an extended, hot halo around NGC~5746.

The spectral analysis of the halo yields a plasma temperature of $kT{\sim}0.6\,\mathrm{keV}$ and an X-ray luminosity of ${\sim}8.8\cdot10^{39}\,\mathrm{erg\,s^{-1}}$ in 0.3--2.0\,keV. The two regions of enhanced emission to the East and West of the disk exhibit slightly higher best-fit temperatures of $kT{\sim}0.7\mbox{--}0.8\,\mathrm{keV}$. Due to the low surface brightness of the diffuse emission, and the dominance of background components, these temperatures are model dependent and subject to large systematic uncertainties. Still, the detection of a hot plasma component is robust.

The morphology of the extraplanar emission is complex. The 0.3--5\,keV EPIC image displays an asymmetric distribution with a preferential extension along the East-West direction. The 0.3--1.2\,keV image shows structured emission in the West, extending away from the minor axis of the galaxy.

We have discussed several possible origin scenarios for the hot halo. Stellar feedback from present or recent star formation is consistent with the observed morphology and temperatures, but is not strictly required by the data. The lack of detectable extraplanar H$\alpha$ and radio continuum emission \citep{rasmussen2009} indicates that an outflow is not currently active. Given the long cooling timescales of hot ($T \gtrsim 10^6\,\mathrm{K}$) plasma, the observed extraplanar emission could be a remnant of past feedback activity, of which the H$\alpha$ and radio tracers have already declined below current sensitivity limits. An estimate of the SFR based on the XRB population suggests that the recent SFR could have been higher than the ${\sim}0.8-1.0\,\mathrm{M_{\odot}\,yr^{-1}}$ \citep{jiang2019} inferred from IR indicators. However, this estimate is accompanied by large systematic uncertainties and should be regarded as a consistency check, rather than a precise measurement. Overall, stellar feedback represents one plausible mechanism that could contribute to the formation and structuring of a hot, quasi-hydrostatic halo. In a system as massive as NGC 5746 ($M_\ast = 1.0\mbox{--} 1.3\cdot10^{11}\,\mathrm{M_{\odot}}$; \citealt{jiang2019}), such a halo can be maintained by the deep gravitational potential and is consistent with observations of other galaxies with comparable mass (e.g., UGCA145 in \citealt{li2017}). Accretion of intergalactic gas may also contribute to the hot gas reservoir. Heating by AGN feedback seems unlikely given the lack of evidence of AGN activity.

A robust measurement of the gas metallicity would provide a key constraint on the origin of the halo emission. Metal-enriched gas would point toward stellar feedback, while a low metallicity would favor accretion scenarios. Unfortunately, the signal-to-noise and spectral resolution of the present data did not allow for a confident measurement of the metallicity.

Overall, our results demonstrate that massive spiral galaxies such as NGC~5746 can host luminous, extended X-ray halos. This result shows that the presence and visibility of extraplanar gas depends not only on current star formation, but also on galaxy mass and the sensitivity limits of current X-ray observatories. Future missions, with improved sensitivity and increased spectral resolution, such as the NewAthena mission, are essential to determine the chemical composition and, ultimately, the origin of the hot halo gas.

\begin{acknowledgements}

This work was supported by the Deutsche Forschungsgemeinschaft through project
SA 2131/18-1 (project no.\ 525647185). It is based on observations obtained with \textit{XMM-Newton}, an ESA science mission with instruments and contributions directly funded by ESA Member States and NASA. The data were obtained from the XMM-Newton Science Archive and processed using the XMM-Newton Science Analysis System (SAS) developed by ESA. Furthermore, this research has made use of XSPEC, developed by Keith Arnaud and maintained by the HEASARC at NASA's Goddard Space Flight Center, and  the SIMBAD database, operated at CDS, Strasbourg, France. 

\end{acknowledgements}

\bibliographystyle{aa}
{\footnotesize
\bibliography{references}}

@article{intr7,
  author =       "K. D. Kuntz and S. L. Snowden",
  journal =      "ApJS",
  volume =       "188",
  pages =       "46",
  year =         "2010"
}

@article{intr8,
  author =       "P. J. Kavanagh and M. Sasaki and D. Breitschwerdt",
  journal =      "A\&A",
  volume =       "637",
  pages =       "A12",
  year =         "2020"
}

@article{wilms,
  author =       "J. Wilms and A. Allen and R. McCray",
  journal =      "ApJ",
  volume =       "542",
  pages =       "914",
  year =         "2000"
}

@article{mineo1,
  author =       "S. Mineo and M. Gilfanov and R. Sunyaev",
  journal =      "MNRAS",
  volume =       "419",
  pages =       "p.2095-21115",
  year =         "2012"
}

@article{mineo2,
  author =       "S. Mineo and M. Gilfanov and R. Sunyaev",
  journal =      "MNRAS",
  volume =       "426",
  pages =       "p.1870–1883",
  year =         "2012"
}

@article{rasmussen2009,
  author =       "Jesper Rasmussen and Jesper Sommer-Larsen and Kristian Pedersen and others",
  journal =      "ApJ",
  volume =       "697",
  pages =       "p.79-93",
  year =         "2009"
}

@article{condon1998,
  author =       "J. J. Condon and W. D. Cotton and E. W. Greisen and others",
  journal =      "AJ",
  volume =       "115",
  pages =       "1693",
  year =         "1998"
}

@article{rand2008,
  author =       "R. J. Rand and R. A. Benjamin",
  journal =      "ApJ",
  volume =       "676",
  pages =       "p.991-1007",
  year =         "2008"
}

@article{barentine2012,
  author =       "J. C. Barentine and J. Kormendy",
  journal =      "ApJ",
  volume =       "754",
  pages =       "140",
  year =         "2012"
}

@article{martig2021,
  author =       "M. Martig and F. Pinna and J. Falcón-Barroso and others",
  journal =      "MNRAS",
  volume =       "508",
  pages =       "p.2458-2478",
  year =         "2021"
}

@article{hopkins2012,
  author =       "Philip F. Hopkins and Eliot Quataert and Norman Murray",
  journal =      "MNRAS",
  volume =       "421",
  pages =       "p.3522-3537",
  year =         "2012"
}

@article{mcquinn2018,
  author =       "Kristen B. W. McQuinn and Evan D. Skillman and Taryn N. Heilman and others",
  journal =      "MNRAS",
  volume =       "477",
  pages =       "p.3164-3177",
  year =         "2018"
}

@article{bianchi2007,
  author =       "S. Bianchi",
  journal =      "A\&A",
  volume =       "471",
  pages =       "765",
  year =         "2007"
}

@article{jiang2019,
  author =       " X. Jiang and J. Li and T. Fang and others",
  journal =      "ApJ",
  volume =       "885",
  pages =       "38",
  year =         "2019"
}

@article{hyperleda,
  author =       "Dimitry Makarov and Philippe Prugniel and Nataliya Terekhova and others",
  journal =      "A\&A",
  volume =       "570",
  pages =       "A13",
  year =         "2014"
}

@book{devaucoleurs1991,
    author = "G. de Vaucoleurs and A. de Vaucoleurs and H. G. Corwin and others",
    title = "Third Reference Catalogue of Bright Galaxies.",
    publisher = "Springer, New York",
    year = "1991"
}

@article{hecate,
  author =       "K. Kovlakas and A. Zezas and J. J. Andrews and others",
  journal =      "MNRAS",
  volume =       "506, Issue 2",
  pages =       "pp.1896-1915",
  year =         "2021"
}

@article{kormendy2019,
  author =       "John Kormendy and Ralf Bender",
  journal =      "ApJ",
  volume =       "872",
  pages =       "106",
  year =         "2019"
}

@article{garcia1993,
  author =       "A. M. Garcia",
  journal =      "A\&ASS",
  volume =       "100",
  pages =       "p.47-90",
  year =         "1993"
}

@article{strickland2004a,
  author =       "David K. Strickland and Timothy M. Heckman and Edward J. M. Colbert and others",
  journal =      "ApJ",
  volume =       "606",
  pages =       "p.829-852",
  year =         "2004"
}

@article{strickland2004b,
  author =       "David K. Strickland and Timothy M. Heckman and Edward J. M. Colbert and others",
  journal =      "AJSS",
  volume =       "151",
  pages =       "p.193-236",
  year =         "2004"
}

@article{tüllmann2006b,
  author =       "R. Tüllmann and D. Breitschwerdt and J. Rossa3 and others",
  journal =      "A\&A",
  volume =       "457",
  pages =       "p.779-785",
  year =         "2006"
}

@article{toft2002,
  author =       "S. Toft and J. Rasmussen and J. Sommer-Larsen and others",
  journal =      "MNRAS",
  volume =       "335",
  pages =       "p.799-806",
  year =         "2002"
}

@article{carillo1999,
  author =       "R. Carrillo and J. Masegosa and D. Dultzin-Hacyan and others",
  journal =      "AA",
  volume =       "35",
  pages =       "187",
  year =         "1999"
}

@article{benson2000,
  author =       "A. J. Benson and R.G. Bower and C. S. Frenk and others",
  journal =      "MNRAS",
  volume =       "314",
  pages =       "557",
  year =         "2000"
}

@article{norman1989,
  author =       "C. A. Norman and S. Ikeuchi",
  journal =      "ApJ",
  volume =       "345",
  pages =       "372",
  year =         "1989"
}

@article{avillez2000,
  author =       "M. A. de Avillez",
  journal =      "MNRAS",
  volume =       "315",
  pages =       "479",
  year =         "2000"
}

@article{bregman1980,
  author =       "J. N. Bregman",
  journal =      "ApJ",
  volume =       "236",
  pages =       "577",
  year =         "1980"
}

@article{shapiro1976,
  author =       "P. R. Shapiro and G. B. Field",
  journal =      "ApJ",
  volume =       "205",
  pages =       "762",
  year =         "1976"
}

@article{breitschwerdt1999,
  author =       "D. Breitschwerdt and T. Schmutzler",
  journal =      "A\&A",
  volume =       "347",
  pages =       "650",
  year =         "1999"
}

@article{breitschwerdt1991,
  author =       "D. Breitschwerdt and H. J. Voelk and J. F. McKenzie",
  journal =      "A\&A",
  volume =       "245",
  pages =       "79",
  year =         "1991"
}

@article{heiles1984,
  author =       "C. Heiles",
  journal =      "ApJS",
  volume =       "55",
  pages =       "585",
  year =         "1984"
}

@article{schulz1992,
  author =       "H. Schulz and G. Wegner",
  journal =      "A\&A",
  volume =       "266",
  pages =       "167",
  year =         "1992"
}

@article{pietsch2000,
  author =       "W. Pietsch and A. Vogler and U. Klein and others",
  journal =      "A\&A",
  volume =       "360",
  pages =       "24",
  year =         "2000"
}

@article{pietsch2001,
  author =       "W. Pietsch and T. P. Roberts and M. Sako and others",
  journal =      "A\&A",
  volume =       "365",
  pages =       "L174",
  year =         "2001"
}

@article{strickland2000,
  author =       "D. K. Strickland and T. M. Heckman and K. A. Weaver and others",
  journal =      "AJ",
  volume =       "120",
  pages =       "2965",
  year =         "2000"
}

@article{tüllmann2000,
  author =       "R. Tüllmann and R.-J. Dettmar",
  journal =      "A\&A",
  volume =       "362",
  pages =       "119",
  year =         "2000"
}

@article{reynolds1984,
  author =       "R. J. Reynolds",
  journal =      "ApJ",
  volume =       "282",
  pages =       "191",
  year =         "1984"
}

@article{wood2004,
  author =       "K. Wood and J. S. Mathis",
  journal =      "MNRAS",
  volume =       "353",
  pages =       "1126",
  year =         "2004"
}

@article{tüllmann2006a,
  author =       "R. Tüllmann and W. Pietsch and J. Rossa and others",
  journal =      "A\&A",
  volume =       "448",
  pages =       "43-75",
  year =         "2006"
}

@article{condon1992,
  author =       "J. J. Condon ",
  journal =      "ARA\&A",
  volume =       "30",
  pages =       "575",
  year =         "1992"
}

@article{wang2001,
  author =       "Q. D. Wang and S. Immler and R. A. M. Walterbos and others",
  journal =      "ApJ",
  volume =       "555",
  pages =       "L99",
  year =         "2001"
}

@article{read2001,
  author =       "A. M. Read and T. J. Ponman",
  journal =      "MNRAS",
  volume =       "328",
  pages =       "127",
  year =         "2001"
}

@article{ranalli2003,
  author =       "P. Ranalli and A. Comastri and G. Setti",
  journal =      "A\&A",
  volume =       "399",
  pages =       "39",
  year =         "2003"
}

@article{ferrara2000a,
  author =       "A. Ferrara and E. Tolstoy",
  journal =      "MNRAS",
  volume =       "313",
  pages =       "291",
  year =         "2000"
}

@article{ferrara2000b,
  author =       "A. Ferrara and M Pettini and Y. Shchekinov",
  journal =      "MNRAS",
  volume =       "319",
  pages =       "539",
  year =         "2000"
}

@article{pedersen2006,
  author =       "K. Pedersen and J. Rasmussen and J. Sommer-Larsen and others",
  journal =      "New Astronomy",
  volume =       "11",
  pages =       "p.465-470",
  year =         "2006"
}

@article{pn,
  author =       "L. Strüder and U. Briel and K. Denner and others",
  journal =      "A\&A",
  volume =       "365",
  pages =       "L18",
  year =         "2001"
}

@article{mos,
  author =       "M. J. L. Turner and A. Abbey and M. Arnaud and others",
  journal =      "A\&A",
  volume =       "365",
  pages =       "L27",
  year =         "2001"
}

@article{HI4PI,
  author =       "N. Ben Bekhti and L. Flöer and R. Keller and others",
  journal =      "A\&A",
  volume =       "594",
  pages =       "A116",
  year =         "2016"
}

@article{apec,
  author =       "R. K. Smith and N. S. Brickhouse and D. A. Liedahl",
  journal =      "ApJ",
  volume =       "556",
  pages =       "L91",
  year =         "2001"
}

@article{henley2013,
  author =       "David B. Henley and Robin L. Shelton",
  journal =      "ApJ",
  volume =       "773",
  pages =       "92",
  year =         "2013"
}

@article{bluem2022,
  author =       "Jesse Bluem and Philip Kaaret and K. D. Kuntzand others",
  journal =      "ApJ",
  volume =       "936",
  pages =       "72",
  year =         "2022"
}

@article{swcx,
  author =       "K. D. Kuntz",
  journal =      "A\&AR",
  volume =       "27",
  pages =       "1",
  year =         "2019"
}

@article{hodges2013,
  author =       "Edmund J. Hodges-Kluck and Joel N. Bregman",
  journal =      "ApJ",
  volume =       "762",
  pages =       "12",
  year =         "2013"
}

@article{gilfanov2004,
  author =       "M. Gilfanov",
  journal =      "MNRAS",
  volume =       "349",
  pages =       "146-168",
  year =         "2004"
}

@article{revnivtsev2008,
  author =       "M. Revnivtsev and E. Churazov and S. Sazonov and others",
  journal =      "A\&A",
  volume =       "490",
  pages =       "p.37-43",
  year =         "2008"
}

@article{tully2007,
  author =       "R. Brent Tully and Luca Rizzi and Edward J. Shaya and others",
  journal =      "AJ",
  volume =       "138",
  pages =       "p.323-331",
  year =         "2009"
}

@article{sorce2014,
  author =       "J. G. Sorce and R. B. Tully and H. M. Courtois and others",
  journal =      "MNRAS",
  volume =       "444",
  pages =       "p.527-541",
  year =         "2014"
}

@article{tully2016,
  author =       "R. Brent Tully and Hélène M. Courtois and Jenny G. Sorce",
  journal =      "AJ",
  volume =       "152, Issue 2",
  pages =       "50",
  year =         "2016"
}

@article{karachentsev2014,
  author =       "I. D. Karachentsev and V. E. Karachentseva and O. G. Nasonova",
  journal =      "Astrophysics",
  volume =       "57, Issue 4",
  pages =       "p.457-472",
  year =         "2014"
}

@article{crain2010,
  author =       "Robert A. Crain and Ian G. McCarthy and Carlos S. Frenk and others",
  journal =      "MNRAS",
  volume =       "407",
  pages =       "p.1403-1422",
  year =         "2010"
}

@article{bogdan2013,
  author =       "Akos Bogdán and William R. Forman and Mark Vogelsberger and others",
  journal =      "ApJ",
  volume =       "772",
  pages =       "97",
  year =         "2013"
}

@article{bregman1994,
  author =       "J. N. Bregman and R. A. Pildis",
  journal =      "ApJ",
  volume =       "420",
  pages =       "570-575",
  year =         "1994"
}

@article{li2017,
  author =       "Jiang-Tao Li and Joel N. Bregman and Q. Daniel Wang and others",
  journal =      "ApJs",
  volume =       "233",
  pages =       "20",
  year =         "2017"
}

@article{white1991,
  author =       "White, Simon D. M. and Frenk, Carlos S.",
  journal =      "ApJ",
  volume =       "379",
  pages =       "52",
  year =         "1991"
}

@article{deluca2004,
  author =       "A. De Luca and S. Molendi",
  journal =      "A\&A",
  volume =       "419",
  pages =       "837-848",
  year =         "2004"
}

@article{li2018,
  author =       "Li, Jiang-Tao and Bregman, Joel N. and Wang, Q. Daniel and others",
  journal =      "ApJ",
  volume =       "855",
  pages =       "L24",
  year =         "2018"
}

@article{li2011,
  author =       "Li, Zhiyuan and Jones, Christine and Forman, William R. and others",
  journal =      "ApJ",
  volume =       "730",
  pages =       "16",
  year =         "2011"
}

@article{li2006,
  author =       "Li, Z. and Wang, Q. D. and Irwin, J. A. and others ",
  journal =      "MNRAS",
  volume =       "371",
  pages =       "147",
  year =         "2006"
}

@article{veilleux2005,
  author =       "Veilleux, Sylvain and Cecil, Gerald and Bland-Hawthorn, Joss and others",
  journal =      "ARA\&A",
  volume =       "43",
  pages =       "769",
  year =         "2005"
}

\end{document}